\documentclass[12pt]{iopart}
\usepackage{latexsym}
\usepackage[T1]{fontenc}
\usepackage[utf8x]{inputenc}
\usepackage{times}
\usepackage{graphicx,amsfonts,amsbsy,amssymb,amsthm}
\usepackage{ bm}
\usepackage{amsopn}
\usepackage{color}
\usepackage{cite}

\def\={\;=\;} \def\+{\,+\,} \def\m{\,-\,}    
     \def\N{\Bbb N} 

\def\tr {\rm tr}

\newtheorem*{prop*}{Proposition}

\newcommand{\beq}{\begin{equation}}

\newcommand{\eeq}{\end{equation}}

\newcommand{\deltaAv}{\bar D}

\def\N{\mathcal{N}}
\def\<{\langle}
\def\>{\rangle}

\theoremstyle{definition}

\theoremstyle{remark}

\begin{document}

\title  {The Statistics of Spectral Shifts due to Finite Rank Perturbations}

\author{Barbara Dietz$^{1}$\footnote{To whom correspondence should be
    addressed (dietz@lzu.edu.cn).}, Holger Schanz$^{2}$ , Uzy Smilansky$^{3}$ and Hans Weidenm\"uller$^{4}$ }

\address { \small{ $^{1}$
 School of Physical Science and Technology,  Lanzhou
 University, Lanzhou,  Gansu 730000,  China \\
 $^{2}$ University of Applied Sciences
 Magdeburg-Stendal, 39114 Magdeburg, Germany\\ 
$^{3}$  Department of Physics of Complex Systems,
Weizmann Institute of Science, Rehovot 76100, Israel\\
  $^{4}$ Max-Planck-Institut $\rm {f\ddot{u}r}$ Kernphysik,  P.O. Box 103980,
69029 Heidelberg, Germany}}

\begin{abstract}
	 This article is dedicated to the following class of problems. Start
         with an $N\times N$ Hermitian matrix randomly picked from a matrix
         ensemble - the reference matrix. Applying a rank-$t$ perturbation to
         it, with $t$ taking the values $1\le t \le N$, we study the 
         difference between the spectra of the perturbed and the reference
         matrices as a function of $t$ and its
         dependence on the underlying universality class of the random matrix
         ensemble. We consider both, the weaker kind of perturbation which
         either permutes or randomizes $t$ \emph{diagonal} elements and a
         stronger perturbation randomizing successively $t$ \emph{rows and
           columns}. In the first case we derive universal expressions in the
         scaled parameter $\tau=t/N$ for the expectation of the variance of
         the spectral shift functions, choosing as random-matrix ensembles
         Dyson's three Gaussian ensembles. In the second case we find an
         additional dependence on the matrix size $N$.
\end{abstract}
\section {Introduction}
An old and intensively investigated subject in matrix theory addresses the
following questions: Given two matrices $A,B$, how does the spectrum of $B$
relate to that of $A$ given some properties of the "perturbation"
$B-A$. Milestones in this field include e.g., Weyl's spectral interlacing
theory \cite {Wey12} which compares the spectra of two matrices that differ in
a rank-one perturbation, Krein's spectral shift theory \cite {Krein} and its
extensions, and the bounds on the spectral difference as discussed e.g., in
the Hoffman-Wielandt theory \cite {Wielandt}. In the present work we address
a subject within the same genre, namely, the differences in the spectral
properties of random $N\times N$ Hermitian matrices $H^{(0)}$ and $H^{(t)}$
with Gaussian distributed entries that differ by a rank-$t$ matrix.  Our goal
is to compute the ensemble averaged difference between their spectra and its
dependence on $t$ and $N$ in terms of Krein's spectral shift. To the best of our knowledge this problem has not been addressed before, although numerous studies exist where the perturbation $V$ in $H^{(t)}=H^{(0)}+tV$  was a fixed random matrix with same symmetry properties as $H^{(0)}$~\cite{Smolyarenko2003,Aleiner1998}. 
It is clear that the results depend on the details
of the chosen perturbation and the matrix ensembles considered. A rank $t$
perturbation which alters all the entries of $t$ rows and their conjugate
columns is expected to have typically a considerably larger effect on the
spectral difference than a rank $t$ perturbation which changes only $t$
diagonal entries.

These two extreme cases are to be discussed for the random Gaussian
orthogonal, unitary and symplectic ensembles \cite{Meh90}. These ensembles play
an important role in the field of Quantum Chaos, as they describe the spectral
properties of typical quantum systems of corresponding universality class with
fully chaotic dynamics in the classical limit
\cite{BGS86,LesHouches89,GMW98,Haa10}.
We denote them by their
"inverse temperature" parameter $\beta
=1,2$ and $4$, respectively. We shall take the limit $N\to\infty $ and $t\to\infty$
keeping their ratio $\tau=t/N$ constant.

The need to consider this problem came from a seemingly unrelated subject,
namely the response of the spectra of quantum graphs to changes of the lengths
of edges while keeping constant the connectivity of the graph and the boundary
conditions at the vertices. Previous theoretical work on quantum graphs
\cite{Berkolaiko,Aizenmann,Schanz} has shown that the exchange of lengths of
pairs of edges results in the interlacing of the spectra in agreement with
Weyl's rule. Namely, the $n$-th eigenvalue of the graph with exchanged lengths
is bounded from below and above by the original eigenvalues with quantum
numbers $n-2$ and $n+2$, respectively.  Such length changes are easy to
realize in experiments on networks of microwave coaxial cables \cite{YBS20} or
optical fibers. In this way, using a given set of cable lengths, one can generate
more spectral data in the finite frequency range where a single transversal
mode is excited. Since the spectral statistics of sufficiently well connected
quantum graphs are reproducible by random matrix theory
(RMT)~\cite{Kottos,Weidenmueller,GA04}, the results obtained in the present
work also offer another possible test concerning the applicability of RMT to
describe spectral properties of quantum graphs.

 The remainder of the introduction is devoted to a description of the various
 kinds of perturbation processes which are to be analyzed. It will be followed
 by the introduction of the spectral measures which are to be used in order to
 quantify the change in the spectra.

\subsection {Random matrices and their perturbations}\label{RMTP}
\noindent To start - a note on notations.  The row and column indices of
vectors and matrices will be denoted by lower-case Greek letters as in e.g.,
$H_{\mu\nu}$ or as in $|\nu\rangle$ which denotes the column vector whose
entries are all $0$ except for a value $1$ at the position $\nu$. The spectral
information will be indexed by lower-case Latin letters. The eigenvector
corresponding to the eigenvalue $\lambda_n$ will be denoted by $| n
\rangle$. The spectra will be ordered monotonically $\lambda_n\le
\lambda_{n+1}$.

We consider $N\times N$ random matrices from the three standard Gaussian
ensembles. The elements are real (GOE, $\beta=1$), complex (GUE, $\beta=2$)
and quaternion real (GSE, $\beta=4$), i.e., they are expressed in terms of
$\beta$ real variables with zero mean and independent normal distributions. We
choose their variance such that the spectra are supported in the interval
$(-1,+1)$ for $N\to\infty$: For the off-diagonal elements $\mu\ne\nu$ the
variance is $\langle|H_{\mu\nu}|^{2}\rangle=(4\beta N)^{-1}$ while Hermiticity
constrains the diagonal elements to be real with variance
$\<|H_{\nu\nu}|^{2}\>=(2\beta N)^{-1}$. Here and in the following $\<\cdot\>$
stands for an average over the random matrix ensemble. With this
normalization, the mean spectral density of the matrices is given by a
semicircle with unit radius. Independent of $\beta$ the corresponding mean
spectral counting function is
\begin{equation}\label{meanN}
\<\N(\lambda)\>=\frac{N}{\pi}\left[\lambda\sqrt{1-\lambda^2}+\frac{\pi}{2}+\arcsin\left(\lambda\right)\right]\qquad(-1\le\lambda\le+1)\,.
\end{equation}
All quantities of interest will be defined such that they are independent of
the matrix size $N$. No unfolding of the spectra will be required, i.e. there is
no correction for the variation of the spectral density along the semicircle.

In the following we introduce rank $t$ perturbations with $t=1,\dots,N$. We
refer to these perturbations as {\em processes} because they can be thought of
as a class of random walks in the space of random matrices. The number of
steps is $t$ (the "time") and we study the evolution of the spectra as the
walk advances from $t=0$ to $t=N$.  The following three perturbation processes
will be discussed.

\paragraph*{(i) Permutation of diagonal matrix elements.} The process
which has the most benign effect on the spectrum consists of a permutation of
$t$ diagonal elements of $H^{(0)}$ while keeping the rest of the matrix
intact. Without loss of generality the permuted entries are chosen as the
first $t$ diagonal elements. Denote the permutation of the first $t$ integers
by $\pi_t$, and assume that it does not have fixed points. We will see below
that under this assumption, different types of permutations have similar
effects.  Let $\pi_t(\kappa)$ and $\pi_t^{-1}(\kappa)$ be the image and
pre-image of $\kappa$, respectively. Then, $\ H^{(t)}$ can be written as
\begin{eqnarray}
\hspace{-10mm} H^{(t)}=H^{(0)}+ \sum_{\kappa=1}^t \left(H^{(0)}_{\pi_t^{-1}(\kappa),\pi_t^{-1}(\kappa)}-H^{(0)}_{\kappa,\kappa}\right)\  |\kappa\rangle\langle \kappa|\ .
\label{tpermutations}
\end{eqnarray}
Clearly we have for the first two traces $\tr [H^{(t)}]=\tr [H^{(0)}]$ and
$\tr [\left (H^{(t)}\right )^2]=\tr [\left (H^{(0)}\right )^2]$. This is a
correlation between the matrices at different values of $t$ that will be
destroyed in the processes $(ii)$, $(iii)$ introduced below.

\paragraph*{(ii) Random replacement of diagonal elements.} Here, the
first $t$ entries are replaced
by new entries $\tilde H_{\kappa,\kappa}$ which are randomly chosen from the
ensemble of which $H^{(0)}$ is a member. The analogue of (\ref
{tpermutations}) is
\begin{equation}
  \hspace{-10mm} \tilde H^{(t)}=H^{(0)}+ \sum_{\kappa=1}^t
  \left(\tilde H_{\kappa,\kappa}-H^{(0)}_{\kappa,\kappa}\right)
  \  |\kappa\rangle\langle \kappa|\  \ .
\label{trandoms}
\end{equation}
In this case only the mean value of the first two traces is constant.
Comparing the spectral changes of $H^{(t)}$ and $\tilde H^{(t)}$ thus opens
the possibility to investigate the effect of the correlations induced in
process {\it(i)} by
requiring the strict invariance of the first two traces.

\paragraph*{(iii) Random replacement of columns and rows.}
The most
severe effect on the spectrum is expected from this process where $t$ columns
and the conjugate rows of $H^{(0)}$ are replaced by the corresponding
elements of another random matrix $H'$ picked from the same ensemble, 
\begin{equation}\label{trandomhans}
  \widetilde H^{(t)}_{\mu\nu}=\left\{
    \begin{array}{ll}
      H^{(0)}_{\mu\nu}\qquad\mbox{for $\mu,\nu\le N-t$}\\[2mm]
      H'_{\mu\nu}\qquad \mbox{otherwise.}
    \end{array}
    \right.
\end{equation}
For $t=N$ the
matrices $H^{(0)}$, $\widetilde H^{(N)}$ have nothing in common except that they belong
to the same random-matrix ensemble.

\paragraph*{} The processes (\emph{i})-(\emph{iii}) certainly do
not cover all possibilities of systematic variations of matrices.  However,
the discussion of these extreme cases provides insight into the range of
spectral variations.  Because of the intrinsic difference between processes
(\emph{i}), (\emph{ii}) and process (\emph{iii}), their treatment requires
disparate methods--perturbative for the first two and field theoretical for
the third type of variation.
 
\subsection {Quantitative  measures for the spectral variation}

A \emph{qualitative} measure for the spectral variation resulting from a rank
$t$ perturbation is given by Weyl's rule \cite{Wey12}:\\
{\em Arrange the spectra $\{ \lambda^{(t)}_{n}\}_{n=1}^N$ in a monotonically
non-decreasing order. Then, the spectra of $H^{(t)}$ and $H^{(0)}$ interlace
so that $\lambda^{(0)}_{n-t}< \lambda^{(t)}_{n}< \lambda^{(0)}_{n+t}$ for
$t<n\le N-t$.}

This relation, however, is not sufficient to distinguish between the
three types of perturbations as they are all of rank $t$. Therefore we employ
more accurate quantitative measures of the spectral difference that will be
introduced in the following two subsections.

\subsubsection{The spectral shift function.}\label{sec:specshift}
\begin{figure}[ht]
\centerline{\includegraphics[width= .6\textwidth]{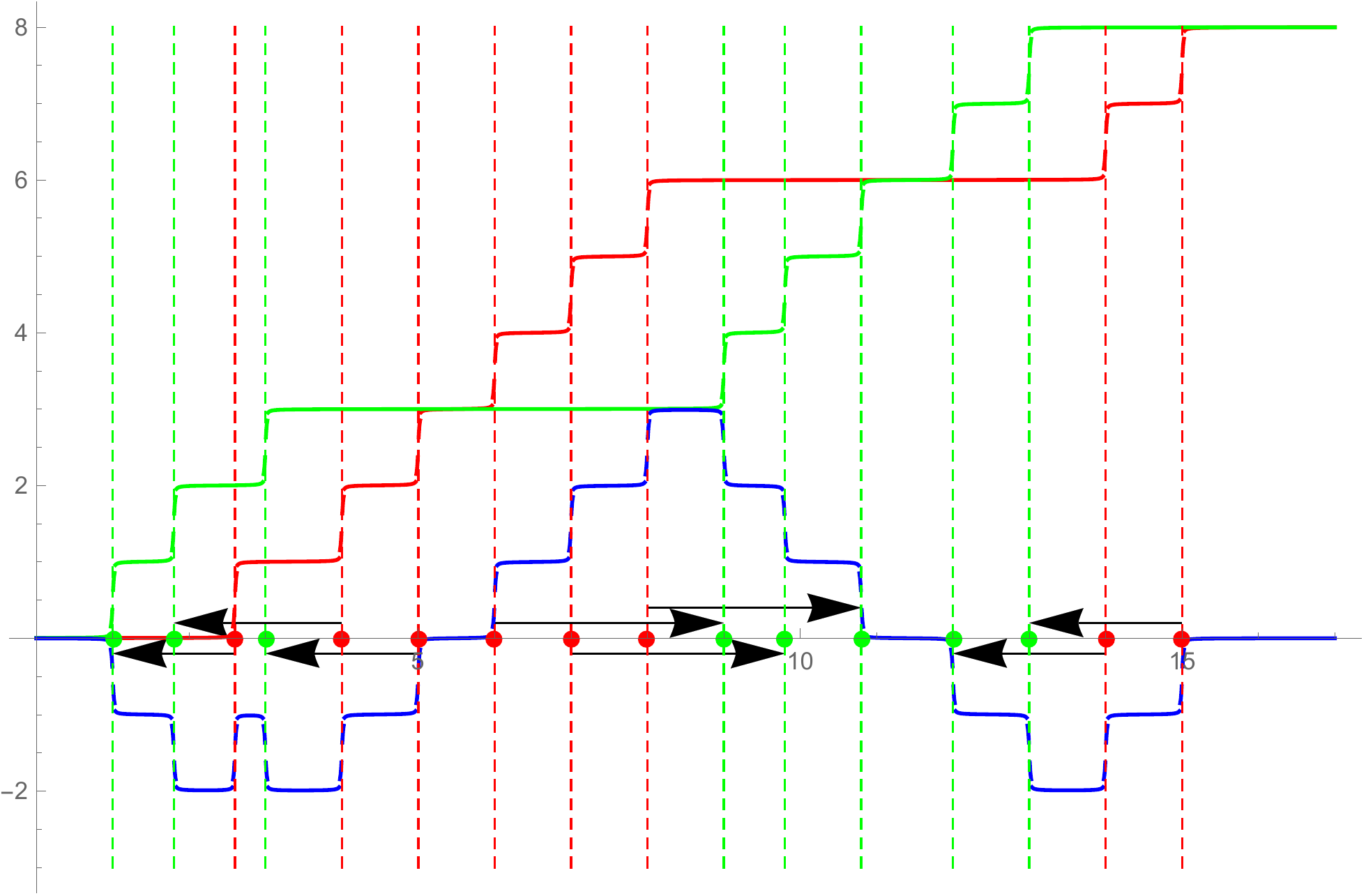}}
\caption{Schematic illustration of Krein's spectral shift
    function (\ref{specshift}). The dots and the vertical dashed lines denote
    the original (green) and the perturbed (red) spectra. The
    corresponding piecewise constant spectral counting functions are shown
    with full lines. They increase by one at each eigenvalue. The blue line is the
    spectral shift function, i.e. the difference of the counting
    functions. The arrows point from the eigenvalue with index $n$ of the perturbed
    spectrum to the corresponding eigenvalue of the original spectrum. Note that the
    absolute value of the spectral shift at any $\lambda$ is
    given by the number of arrows crossing $\lambda$ and its sign is given by
    the orientation of these arrows.
}
\label{deltafig}
\end{figure}

Denote by $\mathcal{N}^{(t)}(\lambda)=\sum_j\theta\left[\lambda-\lambda_j^{(t)}\right]$ 
the spectral counting function, that is the
number of monotonically arranged eigenvalues below a given value $\lambda$. Then
Krein's spectral shift function \cite {Krein} 
\begin{eqnarray}
&&\Delta^{(t)}(\lambda)=\mathcal{N}^{(t)}(\lambda)-\mathcal{N}^{(0)}(\lambda)\label{specshift}\\
&&=\sum_j\left\{\theta\left[\lambda-\lambda_j^{(t)}\right]-\theta\left[\lambda-\lambda_j^{(0)}\right]\right\}
\nonumber\\
&&=\sum_{j=1}^N \Xi_{\lambda^{(0)}_j,\lambda^{(t)}_j }(\lambda)\label{DeltaN} \ ,
\end{eqnarray}
where $\Xi_{a,b}(x)$ is the signed indicator function,
 \begin{eqnarray}
 \Xi_{a,b}(x) =\left \{
 \begin{array}{ll}
 {\rm Sign}(b-a)   &\mbox {{\rm if} \  $\min (a,b) \le x \le \max (a,b) $}\\
 0  &\mbox {${\rm  otherwise}$}
 \end{array}
	 \right . \ \ \ a,b \in \mathbb{R} ,\label{Xi}
	 \end {eqnarray}
is a measure for the difference between the spectra of the original and the
perturbed system. Its definition is illustrated in Fig.~\ref{deltafig}. It is
an integer-valued, piecewise constant function and vanishes outside the
interval which supports the two spectra. The discontinuities occur at
	 points obtained from the union of the two spectra.

	 We define, for a given pair of spectra, the probability to observe a difference $\Delta^{(t)}(\lambda)=r$, if we pick a random value $\lambda$ from their support (which approaches $\lambda\in[-1,+1]$ as $N$ increases, leaving out a negligible and decreasing number of levels), 
\begin{equation}
\hspace{-20mm}
P^{(t)}(r) =  \frac{1}{2}\int_{-1}^{+1} {\rm d}\lambda \
\chi\left(\Delta^{(t)}(\lambda)-r\right);
\ \ \ r\in \mathbb{Z}\,.
\label{valdist}
\end{equation}
The argument of the Kronecker delta function $\chi$ is an integer.
The function equals unity if that integer is zero and vanishes
otherwise. Note that Eq.~(\ref{valdist})
refers to a single matrix and that the integration measure $d\lambda$
  is not modulated by the semicircle density.
%
Clearly $\sum_{r\in \mathbb{Z}}P^{(t)}(r)=1$.
 Furthermore we have
  \begin{eqnarray}
    \sum_{r\in\mathbb{Z}}rP^{(t)}(r)
    &=&\frac{1}{2}\int_{-1}^{+1}d\lambda\,\Delta^{(t)}(\lambda)
    \\&=&-\frac{1}{2}\sum_{n=1}^{N}(\lambda_{n}^{(t)}-\lambda_{n}^{(0)})\label{sumdeltalambda}
    \\&=&-\frac{1}{2}(\tr H^{(t)}-\tr H^{(0)})\,.\label{tracediff}
  \end{eqnarray}  

For a permutation of diagonal matrix elements the trace of the matrix is
conserved and hence, according to Eq.~(\ref{tracediff})
the integral over $\lambda$  of the spectral shift
vanishes identically. While this is not the case for the other
two perturbations that were considered above, it still holds after an ensemble
average.
Therefore we define desymmetrized moments 
\begin{equation}\label{Mt}
M^{(t)}_p= \sum_{r\in \mathbb{Z}} |r|^p\,P^{(t)}(r);\ \ p\in\mathbb{Z}_{+}\,.
\end{equation}
In the sequel we shall deal exclusively with the
first two moments $p=1,2$. As $|r|\le|r|^{2}$ for $r\in\mathbb{Z}$ they obey
$M_{1}^{(t)}\le M_{2}^{(t)}$. Equality holds if the spectral shift takes only
the values $0$, $\pm1$. In particular Weyl's rule implies that this is the
case for $t=1$.
In analogy to Eq.~(\ref{sumdeltalambda}), for the first moment $M_{1}^{(t)}$ a
representation in terms of eigenvalue shifts can be found
  \begin{eqnarray}\label{M1.1}
    M_{1}^{(t)}&=&\frac{1}{2}\int_{-1}^{+1}d\lambda\,|\Delta^{(t)}(\lambda)|
    \\&=&\frac{1}{2}\sum_{n=1}^{N}|\lambda_{n}^{(t)}-\lambda_{n}^{(0)}|\,.
    \label{M1}
  \end{eqnarray}
  This can be understood from Fig.~\ref{deltafig}. All arrows from
  $\lambda_{n}^{(t)}$ to $\lambda_{n}^{(0)}$ crossing a given point $\lambda$ have the
  same direction because the spectra are ordered.
  Thus
  $|\Delta^{(t)}(\lambda)|$ is just the number of arrows crossing
  $\lambda$ implying that the contribution of each arrow to the integral in (\ref{M1.1}) 
  corresponds to its length $|\lambda_{n}^{(t)}-\lambda_{n}^{(0)}|$.
  Similarly, based on Fig.~\ref{deltafig}, it is
  possible to derive an expression for the second moment, which is more
  involved and clarifies in more detail the relation between $M_{1}^{(t)}$ and
  $M_{2}^{(t)}$.


\subsubsection{The distribution of single eigenvalue shifts.}\label{sec:evshifts}
\def\s{s} We consider the shift of individual eigenvalues 
\begin{equation}\label{s}
  \s_{n}^{(t)}=\frac{\lambda_n^{(t)}-\lambda_n^{(0)}}{\deltaAv}
\end{equation}
scaled by the global mean level spacing
\begin{equation}\label{mspc}
  \deltaAv=2/N
\end{equation}
in order to remove the dependence on the matrix size $N$. 
For a pair of spectra we define the probability distribution
\begin{equation}
p(s; t)=\frac{1}{N}\sum_{n=1}^{N}\delta \left(\s-\s_{n}^{(t)}\right)\,.
\label{pkxt}
\end{equation}
In terms of Fig.~\ref{deltafig} $p(s; t)$ can be interpreted as the
distribution of directed arrow lengths. As in the distribution of the spectral
shift function we use in Eq.~(\ref{pkxt}) the whole spectrum of a single
realization without taking into account the dependence of the mean level
spacing on $\lambda$ and, in the present section, do not consider ensemble averages. 

The desymmetrized moments of the distribution are denoted by 
\begin{eqnarray}
  m_{p}^{(t)}&=&\int_{-\infty}^{+\infty}ds\,|s|^{p}p(s;t)
  \\&=&\frac{1}{N}\sum_{n=1}^{N}|\s_n^{(t)}|^{p}\,.
  \label{mp}
\end{eqnarray}
Again we shall study explicitly only the lowest two moments $p=1,2$.
Comparing Eqs.~(\ref{M1}), (\ref{mp}) and using Eq.~(\ref{mspc}) we have for
$p=1$
\begin{eqnarray}\label{m1M1}
  m_{1}^{(t)}&=&M_{1}^{(t)}\,.
\end{eqnarray}

A rough estimate for $m_{p}^{(t)}$ can be obtained on the basis of the {\it Hoffman-Wielandt
  inequality} \cite{Wielandt}:\\
{\em Let $A$ and $B$ be two Hermitian matrices of dimension $N$ and the
spectra $\lambda(A)$ and $\lambda (A+B)$  be arranged in an increasing order.
Then, for any real $p$ with $1\le p \le \infty $
\begin{equation}\label{WHie}
||{\bf \lambda} (A+B)-{\bf \lambda} (A)||_{p} \le ||{\bf \lambda} (B)||_{p}
\end{equation}
where the p-norm of a vector $x$ is $||x||_{p}=(\sum_{n}|x_{n}|^{p})^{1/p}$.
}

In our case $A=H^{(0)}$ and $B=H^{(t)}-H^{(0)}$. Clearly, the application of
this inequality is simplest when studying the processes \emph{(i)} and
\emph{(ii)} where the perturbation is a diagonal matrix with $t$ non-zero
entries. 
Below, in Section \ref{processII}, we shall apply the Hoffman-Wielandt estimate
in the discussion of the random replacement process.  

\section {The spectral variation for modified diagonal elements}\label{diagPerturb}
We continue by addressing the effect of the replacement of $t$ diagonal
elements by newly generated random entries, or the permutation of $t$ diagonal
entries, on the spectrum of a Gaussian random matrix. In the
first three subsections we derive analytical expressions for the spectral
measures. The comparison with numerical simulations and the discussion of
the results will be deferred to Section \ref{numerics}.

\subsection{First-order perturbation theory}
We generally consider ensemble averages of the spectral measures introduced in
the previous section, $M^{(t)}_{1,2}$, $m^{(t)}_{1,2}$, $p(s,t)$ but omit the
explicit notation $\<\cdot\>$ for ensemble averages over these quantities.  We
start with $p(s;t)$ and show that in a certain range of $t$ it has a Gaussian
distribution with known variance $m_{2}^{(t)}$. This allows the computation of
the two lowest moments $M_{1,2}^{(t)}$ of the spectral shift function. The
crucial observation is, that for the Gaussian
matrix ensembles, first order perturbation theory for the spectral differences
$(\lambda_{n}^{(t)}-\lambda_{n}^{(0)})$ can be justified in the limit of large
$N$ and $t\ll N$.  Explicitly, first-order perturbation theory entails the result that the change of eigenvectors induced by the perturbation is of second order only, and accordingly yields for the
permutation process
\begin{equation}
  \lambda_n^{(t)}-\lambda_n^{(0)} \ =  \ \sum_{\kappa= 1}^t (H_{\pi_t^{-1}(\kappa),\pi_t^{-1}(\kappa)}-H_{\kappa,\kappa})\  |\langle n|\kappa\rangle|^2\  ,
 \label{trandoms1st1}
 \end{equation}
	and for the random replacement process
 \begin{equation}
 \tilde \lambda_n^{(t)}-\lambda_n^{(0)} \  =   \ \sum_{\kappa=1}^t (\tilde  H_{\kappa,\kappa}-H_{\kappa,\kappa})\  |\langle n|\kappa\rangle|^2 \ ,
\label{trandoms1st2}
\end{equation}
where $|n\rangle$ and $\lambda_n^{(0)}$ are the eigenvector and the eigenvalue
of the unperturbed matrix $H$. To estimate the magnitude of these terms,
recall the scaling of the matrix elements introduced in Section 1.1, namely
$\<H_{\mu,\nu}^{2}\>\sim N^{-1}$. Thus, the terms
in parantheses are of order $N^{-1/2}$. Furthermore, because
of the uniform distribution of the eigenvectors and their normalization, 
the expansion coefficients
$|\<n|\kappa\>|^{2}$ are of order $N^{-1}$. Therefore the $t$ summands in
(\ref{trandoms1st1}), (\ref{trandoms1st2}) are identically distributed random
variables with zero mean and a finite variance $\sim N^{-3}$. We will argue
below that they can be considered as statistically independent. Hence the
variance grows linearly in $t$ and the sums are of order $\sqrt{\tau}/N$, where
$0\le\tau\le1$ is defined as $\tau=t/N$. Comparing this to the mean level
spacing $\deltaAv\sim N^{-1}$ we conclude that
$\lambda_{n}^{(t)}-\lambda_{n}^{(0)}\sim\sqrt{\tau}\,\deltaAv$, i.e. for $t\ll
N$ ($\tau\ll1$) the eigenvalue shifts are much smaller than the mean level
spacing and perturbation theory is applicable. On the other hand, for $t\sim
N$ ($\tau\sim 1$) the eigenvalue shifts are of the order of the mean
spacing. Then effects beyond first order perturbation theory such as avoided
level crossings will become important.

In order to substantiate the above estimates we need to determine the
statistics of the summands in (\ref{trandoms1st1}),
(\ref{trandoms1st2}). First we note that the unit-norm eigenvectors of Gaussian random matrices can be converted under the transformation associated with their universality class into any arbitrary unit-norm vector. Consequently, the only requirement on the expansion coefficients $x^{(n)}_{\kappa} =N|\langle n|\kappa\rangle|^2$ is that their normalization $\sum_\kappa N\langle n|\kappa\rangle|^2 =\sum_n N\langle n|\kappa\rangle|^2 =1$ remains unchanged. Accordingly, for large $N$ the $x^{(n)}_{\kappa}$ are statistically independent and follow the Porter-Thomas distribution \cite{Haa10,Alonso2016,PKP19}
\begin{eqnarray}\label{PTdist}
p^{PT}_{\beta}(x)&=&\frac{(\beta/2)^{\beta/2}}{\Gamma(\beta/2)}\,x^{\beta/2-1}\exp(-\beta x/2).
\end{eqnarray}
Thus, $\langle x\rangle=1$ and $\langle x^2\rangle = (1+\frac{2}{\beta})$
for the three values of $\beta$. The method used in Ref.~\cite{Haa10} to derive Eq.~(\ref{PTdist}) can be readily generalized to prove that the joint probablity distribution of $t$ expansion coefficients of a given eigenvector can be approximately factorized with an error of order $\frac{t}{N}$. Second, the variances of the diagonal
elements are $\<H_{\mu,\mu}^{2}\>=(2\beta N)^{-1}$, see Section \ref{RMTP}.
Now each term in the sums (\ref{trandoms1st1}), (\ref{trandoms1st2}) is of the
form $H_{\mu,\mu}\ |\langle n|\kappa\rangle|^2$ (where $\mu$ not necessarily
equals $\kappa$) or $\tilde H_{\kappa,\kappa} \ |\langle
n|\kappa\rangle|^2$. The computation is simple for the latter expression since
$\tilde H$ and the eigenvectors of $H$ are statistically independent from each other. We find
 \begin{equation}
  \hspace {-10mm} \left \langle \tilde  H_{\kappa,\kappa} \  |\langle
  n|\kappa\rangle|^2 \right \rangle = 0 \ \ {\rm and} \ \    \left
  \langle \left (\tilde  H_{\kappa,\kappa} \  |\langle
  n|\kappa\rangle|^2\right )^2\right \rangle =
  \frac{c_{\beta}}{2N^3}
   \label{sigma1}
 \end{equation}
 where the constant
 \begin{equation}
   c_{\beta}=(2+\beta)/\beta^{2}
 \end{equation}
 takes the values 3 for GOE, 1 for GUE and 3/8 for GSE, respectively.

The computation of the mean and variance of $H_{\mu,\mu}\ |\langle
n|\kappa\rangle|^2$ is more involved since the eigenvectors of $H$ are
statistically correlated with the matrix elements. However, we can use the
spectral decomposition of $H_{\kappa,\kappa}$ yielding
\begin{equation}
H_{\mu,\mu}\  |\langle n|\kappa\rangle|^2  = \sum_{r=1}^N \lambda_r \ |\langle r|\mu\rangle|^2\ |\langle n|\kappa\rangle|^2\ .
\end{equation}
The spectral parameters which appear in this expression are statistically
independent for the Gaussian ensembles.
In particular, $\langle \lambda_r \rangle =0$ and $ \left \langle
H_{\mu,\mu}\ |\langle n|\kappa\rangle|^2 \right \rangle =0$. Using the fact
that $\left \langle (\tr H)^2\right \rangle = \frac{1}{2\beta}$, and $\left
\langle \tr H^2 \right \rangle = \frac{N}{4} +\frac{1}{4\beta}(2-\beta)$ we
get for $\mu\ne \kappa$,
\begin {eqnarray}
\left \langle \left (H_{\mu,\mu} \  |\langle n|\kappa\rangle|^2\right
)^2\right \rangle =
\frac{c_{\beta}}{2 N^3}\ .
   \label{sigma2}
\end{eqnarray}
This expression coincides with (\ref {sigma1}) as expected from the
statistical independence of the different components of the eigenvectors.  A
slightly more involved computation for the case $\mu=\kappa$ gives,
\begin {eqnarray}
\left \langle \left (H_{\kappa,\kappa} \  |\langle n|\kappa\rangle|^2\right
)^2\right \rangle =
\frac{c_{\beta}}{2N^3}
\left (1+\mathcal{O}\left (\frac{1}{\beta N}\right ) \right )\,,
   \label{sigma3}
\end{eqnarray}
where the derivation of the higher-order term is quite lengthy but straight forward. Comparing (\ref
{sigma2}) and (\ref{sigma3}) with (\ref {sigma1}), we conclude that in the limit of
large $N$, one can consider the eigenvectors and the diagonal matrix elements
as statistically independent variables.
In the following we apply these intermediate results to the two perturbation processes
which are of interest in this paper.

\subsection{Random replacements of diagonal matrix elements}\label{processII}
For $t<N$, the right hand side of Eq.~(\ref {trandoms1st2}) can be approximated as a sum of $t$ independent and
identically distributed terms with the known variance $\sigma^2= c_{\beta}/N^{3}$.
Hence, the variance of  $\lambda_{n}^{(t)}-\lambda_{n}^{(0)}$, averaged over
$n$ and the ensemble, is 
\begin{eqnarray}
  \left\langle N^{-1}\sum_{n=1}^{N}\left|\tilde\lambda_n^{(t)}-\lambda_n^{(0)}
  \right|^{2}\right
  \rangle
  &=&t\,\sigma^{2}
\end{eqnarray}
and thus
\begin{equation}\label{m2}
  m_{2}^{(t)}=\frac{c_{\beta}}{4}\tau\,.
\end{equation}
Furthermore, for large $t$, the arguments presented in the previous section allow to apply the Central Limit Theorem to the right hand side of Eq.~(\ref
{trandoms1st2}). Hence, the quantities $(\tilde \lambda_n^{(t)}-\lambda_n^{(0)})/t$ 
distribute normally with zero mean and with variance $\sigma^2/t$, that is, to
leading order in $1/\sqrt{t}$ the ensemble average of the distribution (\ref {pkxt})
of scaled level shifts is Gaussian and independent of
$n$,
\begin{equation}
  p(s,t)=\sqrt{\frac{2}{\pi c_{\beta}\tau}}
      \exp \left(-\frac{2s^2}{c_{\beta}\tau}\right)\,,
\label{pnx}
\end{equation}
yielding the mean value of $|s|$
\begin{equation}\label{m1}
m_{1}^{(t)}\equiv M_{1}^{(t)}=\sqrt{\frac{c_{\beta}}{2\pi}}\tau^{\frac{1}{2}}
=\sqrt{\frac{2}{\pi}m_{2}^{(t)}}\,.
\end{equation}
Note that for $\tau\ll 1$ the typical eigenvalue shift is much smaller than the
mean level spacing and it is expected that the spectral shift function
is almost always bounded in absolute value by $1$. Hence we have in this
case
\begin{equation}\label{M2rand}
  M_{2}^{(t)}\approx M_{1}^{(t)}=
  \sqrt{\frac{c_{\beta}}{2\pi}}\tau^{\frac{1}{2}}\qquad(\tau\ll 1)\,. 
\end{equation}
We will discuss the accuracy of this result in
Sec.~\ref{numerics} and its implications in the Conclusions Sec.~\ref{conclusions}.

It is interesting to compare our results with the bounds provided by the
Hoffman-Wielandt inequality (\ref{WHie}). The l.h.s. contains a sum over a power of
the $N$ eigenvalue shifts, while on the r.h.s. the $t$ diagonal elements of the
perturbation are summed up. We find   
\begin{equation}
  N m_{p}^{(t)}
  \le t\<|H_{\kappa,\kappa}^{(t)}-H_{\kappa,\kappa}^{(0)}|^{p}\>\,.
\end{equation}
The perturbation matrix has Gaussian
distributed diagonal entries $\tilde H_{\kappa,\kappa}-H_{\kappa,\kappa}$ with
variance $(\beta N)^{-1}$ and mean absolute value $\sqrt{2/N\pi\beta}$. Thus
\begin{eqnarray}\label{WHm1}
  m_{1}^{(t)}&\le&\sqrt{\frac{N}{2\pi\beta}}\,\tau
  \\m_{2}^{(t)}&\le&\frac{N\tau}{4\beta}\label{WHm2}\,.
\end{eqnarray}
The variance $m_{2}^{(t)}$ as given by (\ref{m2}) is compatible with
(\ref{WHm2}) for $N\to\infty$.  On the other hand, after substitution of
(\ref{m1}), equality holds in (\ref{WHm1}) for $t=\beta c_{\beta}$ which is
$t=3$ for $\beta=1$, $t=2$ for $\beta=2$ and $t=3/2$ for $\beta=4$. For
smaller $t$ the inequality contradicts our result (\ref{m1}) for
$m_{1}^{(t)}$ which is valid for $t\to\infty$.

\subsection {Permuting diagonal matrix elements}\label{processI}
There are two features which render the treatment of the permutation process
more involved than the study of the randomization 
discussed above. The first one is that under
permutations, both ${\rm tr} H^{(t)} $ and ${\rm tr} \left (H^{(t)}\right )^2
$ do not vary with $t$. This invariance might impose constraints on the
spectral shift which do not exist in the treatment of the randomization process and
could restrict the applicability of first-order perturbation which
was the prerequisite for the derivations in the previous case. To check this
possibility, recall the first order expression for the eigenvalue difference
(\ref {trandoms1st1}):
\begin{equation}
  \lambda_n^{(t)}-\lambda_n^{(0)} \ =  \ \sum_{\kappa=1}^t (H_{\pi_t^{-1}(\kappa),\pi_t^{-1}(\kappa)}-H_{\kappa,\kappa})\  |\langle n|\kappa\rangle|^2\  ,
 \label{trandoms1st1r}
 \end{equation}
Then,
\begin{eqnarray}
\hspace{-20mm}
 {\rm tr} H^{(t)}-{\rm tr} H^{(0)}&=&\sum_{n=1}^N
 (\lambda_n^{(t)}-\lambda_n^{(0)}) \nonumber\\
 &=&  \ \sum_{\kappa=1}^t (H_{\pi_t^{-1}(\kappa),\pi_t^{-1}(\kappa)}-H_{\kappa,\kappa})\ \sum_{n=1}^N |\langle n|\kappa\rangle|^2\  =0 ,
 \label{invariance1 }
 \end{eqnarray}
since the last sum over $n$ gives $1$. Thus, first order perturbation theory
preserves ${\rm tr} H^{(t)}$. However perturbation theory fails to comply with
the demand ${\rm tr} \left (H^{(t)}\right )^2 -{\rm tr} \left (H^{(0)}\right
)^2 =0$.  To leading order
\begin{eqnarray}
\hspace{-20mm}
 {\rm tr} (H^{(t)} )^2-{\rm tr} ( H^{(0)} )^2 &=&\sum_{n=1}^N ((\lambda_n^{(t)})^2-(\lambda_n^{(0)})^2) \ \approx 2 \sum_{n=1}^N \lambda_n^{(0)}((\lambda_n^{(t)}) -(\lambda_n^{(0)}) ) \nonumber \\
 &=& 2 \sum_{\kappa=1}^t (H_{\pi_t^{-1}(\kappa),\pi_t^{-1}(\kappa)}-H_{\kappa,\kappa})\ \sum_{n=1}^N \lambda_n^{(0)}|\langle n|\kappa\rangle|^2     \nonumber \\
 &=&  2 \sum_{\kappa=1}^t (H_{\pi_t^{-1}(\kappa),\pi_t^{-1}(\kappa)}-H_{\kappa,\kappa})\ H_{\kappa,\kappa}  \nonumber \\
 &=&    \sum_{\kappa=1}^t \left [2  H_{\pi_t^{-1}(\kappa),\pi_t^{-1}(\kappa)} H_{\kappa,\kappa} -H_{\kappa,\kappa} ^2 -H_{\pi_t^{-1}(\kappa),\pi_t^{-1}(\kappa)}^2\right ] \nonumber \\
 &=&    -\sum_{\kappa=1}^t \left [  H_{\pi_t^{-1}(\kappa),\pi_t^{-1}(\kappa)}- H_{\kappa,\kappa}  \right ]^2 \ .
 \label{invariance2}
 \end{eqnarray}
\noindent This term is of order $\frac{t}{N}$, and it can be neglected
in comparison to the expected value of $\left \langle {\rm tr} H^2\right \rangle =
\frac{N}{4} +\frac{1}{4\beta}(2-\beta)$.  Therefore, first-order perturbation
theory is not ruled out by the invariance of the trace of $H^{2}$ under a
permutation of diagonal elements.

The second feature which complicates the study of the permutation process
is that the terms in the sum (\ref{trandoms1st1r}) are correlated,
and the correlations depend on the decomposition of $\pi_t$ into irreducible
cycles. To illustrate the difficulties which arise due to these correlations, we
shall consider two examples. In the first, which applies to even $t$ only,
$\pi_t$ is chosen as a product of cycles of length 2. It can be always
presented as $\pi_t = (1,2)(3,4)...(t'-1,t')$ where $t'=\frac{t}{2}$.  Then with Eq.~(\ref{trandoms1st1}),
\begin{equation}
\hspace {-10mm}
  \lambda_n^{(t)}-\lambda_n^{(0)}  =   \sum_{\kappa=1}^{t'}  (H_{2\kappa-1,2\kappa-1}-H_{2\kappa ,2\kappa}) ( |\langle n|2\kappa\rangle|^2 -|\langle n|2\kappa-1\rangle|^2)
  \label{sumtwo}
\end{equation}
 The $t'$ terms in the sum are uniformly distributed with zero mean and they
 are statistically independent since each term refers to a unique
 $2$-cycle. The Porter-Thomas distribution gives $\left \langle ( |\langle
 n|2\kappa\rangle|^2 -|\langle n|2\kappa-1\rangle|^2)^2\right \rangle
 =\frac{1}{N^2} \frac{4}{\beta}$. Thus, with the definition
 \begin{equation}\label{cbp}
 {c_{\beta}}'=2/\beta^2
 \end{equation}
 the variance of each term in (\ref{sumtwo}) is
 $\sigma^2=2{c_{\beta}}'/N^{3}$. A factor 2 has been introduced here because the
 sum (\ref{sumtwo}) contains only $t'=t/2$ terms. After this modification we
 can follow the same steps as previously for randomized diagonal elements and
 recover Eqs.~(\ref{m2}), (\ref{pnx}) and (\ref{m1}) with $c_{\beta}$ replaced
 by ${c_{\beta}}'$ .
 
The second example to be considered is a cyclic permutation $\pi_t =
(1,2,\cdots, t)$. Writing $ H_{t+1,t+1}=H_{1,1} $ and $\langle
n|0\rangle=\langle n|t\rangle $, we can write with Eq.~(\ref{trandoms1st1}) the spectral difference
as
\begin{eqnarray}
  \lambda_n^{(t)}-\lambda_n^{(0)} \ &=&  \ \sum_{\kappa=1}^t (H_{\kappa+1,\kappa+1}-H_{ \kappa , \kappa })\  |\langle n|\kappa\rangle|^2\     \\
  &=&\sum_{\kappa=1}^t H_{ \kappa , \kappa }\ ( |\langle n|\kappa-1\rangle|^2-  |\langle n|\kappa\rangle|^2 )
 \label{ex2}
 \end{eqnarray}
Clearly, every term in the sum is correlated with its predecessor.  Therefore
the Central Limit Theorem cannot be used as was previously done in the
derivation of Eqs.~(\ref{pnx}) and~(\ref{m1}) and will be replaced by the Martingale
Central Limit Theorem. Calling the sum in (\ref{ex2}) $S_{t}$
and removing the first term $\kappa =1$ we get
\begin{equation}
\hat S_t=\sum_{\kappa=2}^t H_{ \kappa , \kappa }\ ( |\langle n|\kappa-1\rangle|^2-  |\langle n|\kappa\rangle|^2 )
\label{martingale1}
\end{equation}
which has the following properties making  it a martingale:
\begin{itemize}
\item[-] $\left \langle \hat S_t \right \rangle  = 0$\ .
\item[-] The expectation value of $\hat S_{t+1}$ conditioned on $\hat
  S_1,\cdots,\hat S_t$ having \emph{specific values} equals $\hat S_t$.
\end{itemize}
The second property follows from the fact that
\begin{equation}
\hat S_{t+1}=\hat S_t + H_{ t+1 , t+1 }\ ( |\langle n|t \rangle|^2-  |\langle n| t+1\rangle|^2 )
\end{equation}
where $H_{t+1,t+1}$ distributes independently of $( |\langle n|t \rangle|^2-
|\langle n| t+1\rangle|^2 )$ in the large $N$ limit with $\left\langle
H_{t+1,t+1} \right \rangle =0$.

Moreover, since  the distribution of both the diagonal matrix elements of $H$
and the squared amplitudes of the eigenvectors decay exponentially fast, the
Lindeberg condition \cite {Brown} is satisfied. Therefore the $\hat S_t$
distribute normally with variance $(t-1)\sigma^2$ where $\sigma^{2}=c_\beta'/N^{3}$ is the
variance of any single summation term in Eq.~(\ref{martingale1}). 
For large $t$ this result cannot be substantially changed by the missing term
$t=1$ such that $S_t=\lambda_{n}^{(t)}-\lambda_{n}^{(0)}$ also distributes
normally with variance $t\sigma^{2}$. We are grateful to Jonathan Breuer for
providing us with a formal proof of this claim \cite{Breuer}.

The result above shows that the mean spectral measures for the two extreme
permutations involving the smallest and the largest cycle length,
respectively, are identical. However, general permutations can
display a more complicated cycle structure than the ones discussed above. A
general permutation can be written as a product of $k$ cycles of lengths
$(c_1,\cdots,c_k)$, and $ \sum_k c_k =t$. The sum (\ref{trandoms1st1}) can be
divided into the contribution of statistically independent sums over cycles of
period $c_j$. The number of permutations of $t$ numbers, with $k$ cycles is
given by the Stirling number of the first kind
$\scriptsize{\left[\begin{array}{c}t\\k\end{array}\right]}$ \cite {handbook}.
Considering a random permutation, it is known \cite {Louchard} that for large
t, the distribution of the number of cycles $k$ is approximately normal with
mean $[\log t +\gamma +\frac{1}{2t} +\mathcal{O}(\frac{1}{t^2})]$ and variance
$[\log t -\frac{\pi^2}{6} +\gamma +\frac{3}{2t} +\mathcal{O}(\frac{1}{t^2})]$,
where $\gamma$ is the Euler constant.  Hence, most cycles are long, with $c_j$
of order $t/\log t$.  Therefore, by using the previous result, their
contribution will be normally distributed with variance $c_j \sigma^2$ and
their sum will be distributed normally with variance $t\sigma^2$.

We end this section by summarizing the results obtained for the spectral
measures in the case of permutations of diagonal matrix elements:
\begin{equation}\label{m2p}
  m_{2}^{(t)}\approx \frac{{c_{\beta}}'}{4}\tau\qquad(\tau\ll1)\,,
\end{equation}
\begin{equation}\label{M2p}
  M_{2}^{(t)}\approx M_{1}^{(t)}=m_{1}^{(t)}\approx 
  \sqrt{\frac{{c_{\beta}}'}{2\pi}}\tau^{\frac{1}{2}}\qquad(t\gg1,\ \tau\ll 1)\,. 
\end{equation}

\subsection {Simulations and error estimates}\label{numerics}
\begin{figure}
\centerline{
  \includegraphics[width=0.33\textwidth]{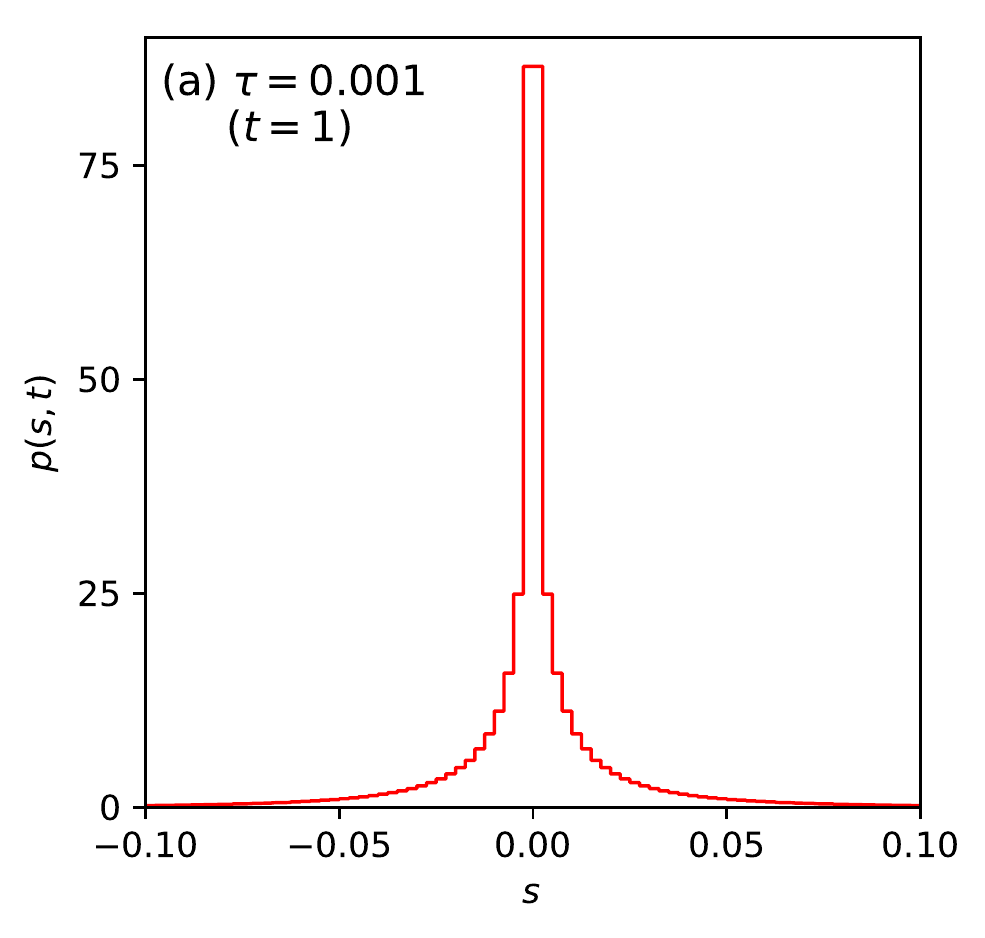}
  \includegraphics[width=0.33\textwidth]{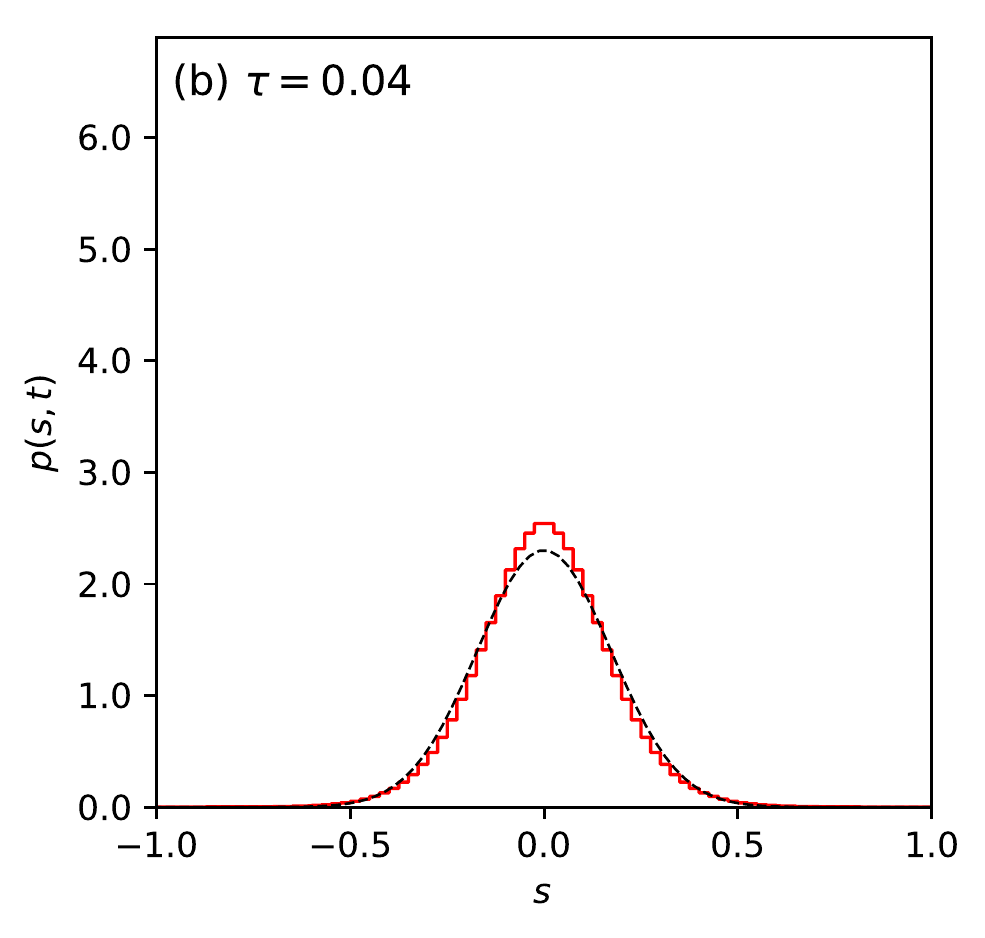}
  \includegraphics[width=0.33\textwidth]{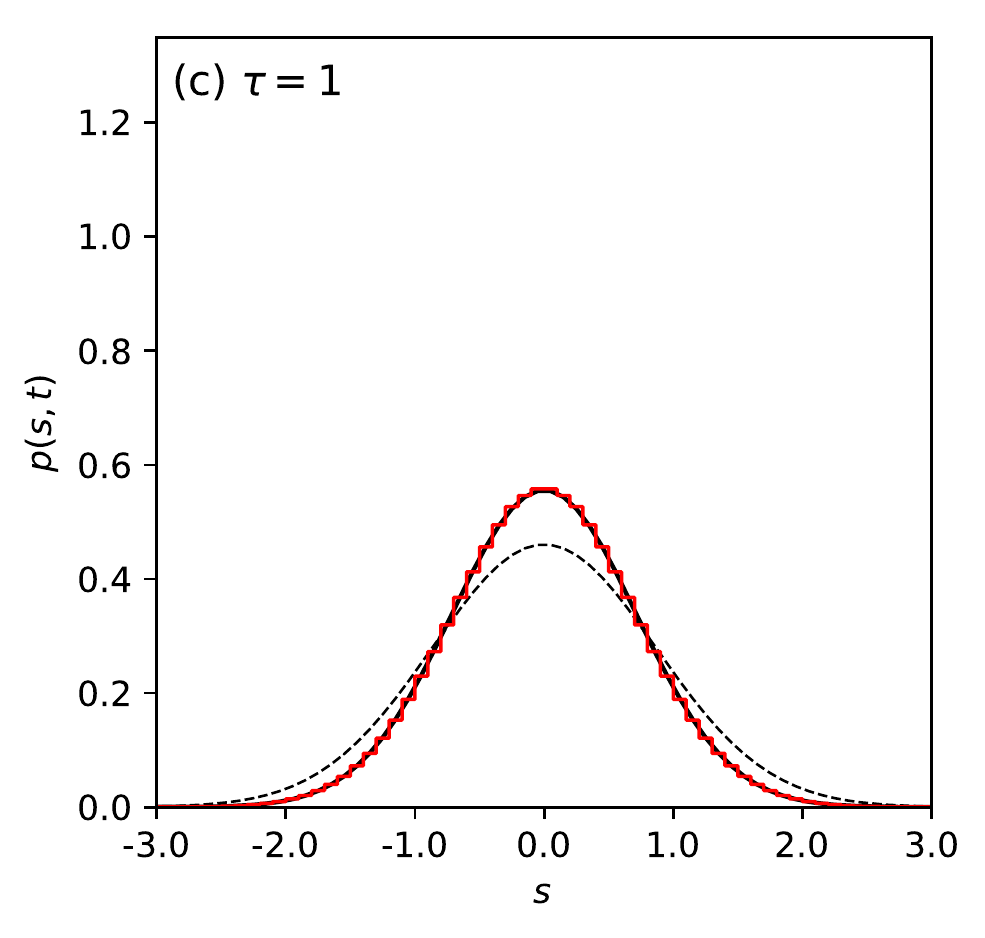}
}
\centerline{
  \includegraphics[width=0.33\textwidth]{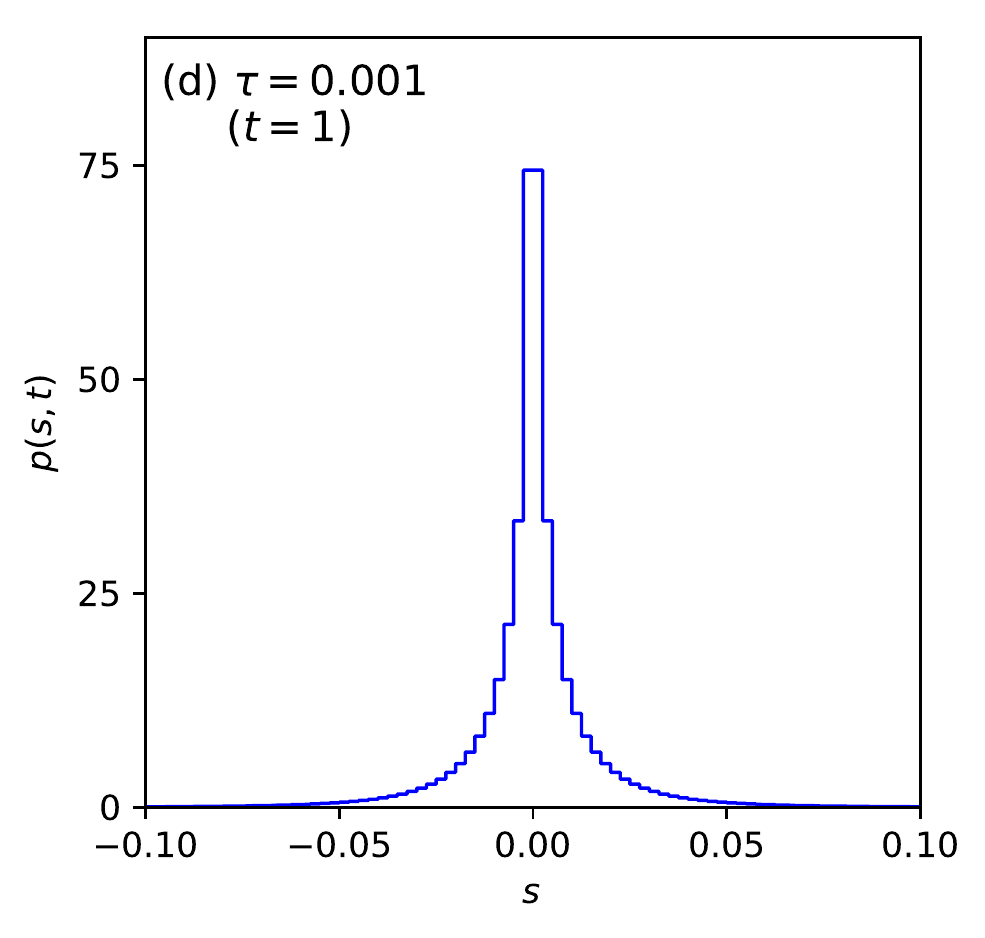}
  \includegraphics[width=0.33\textwidth]{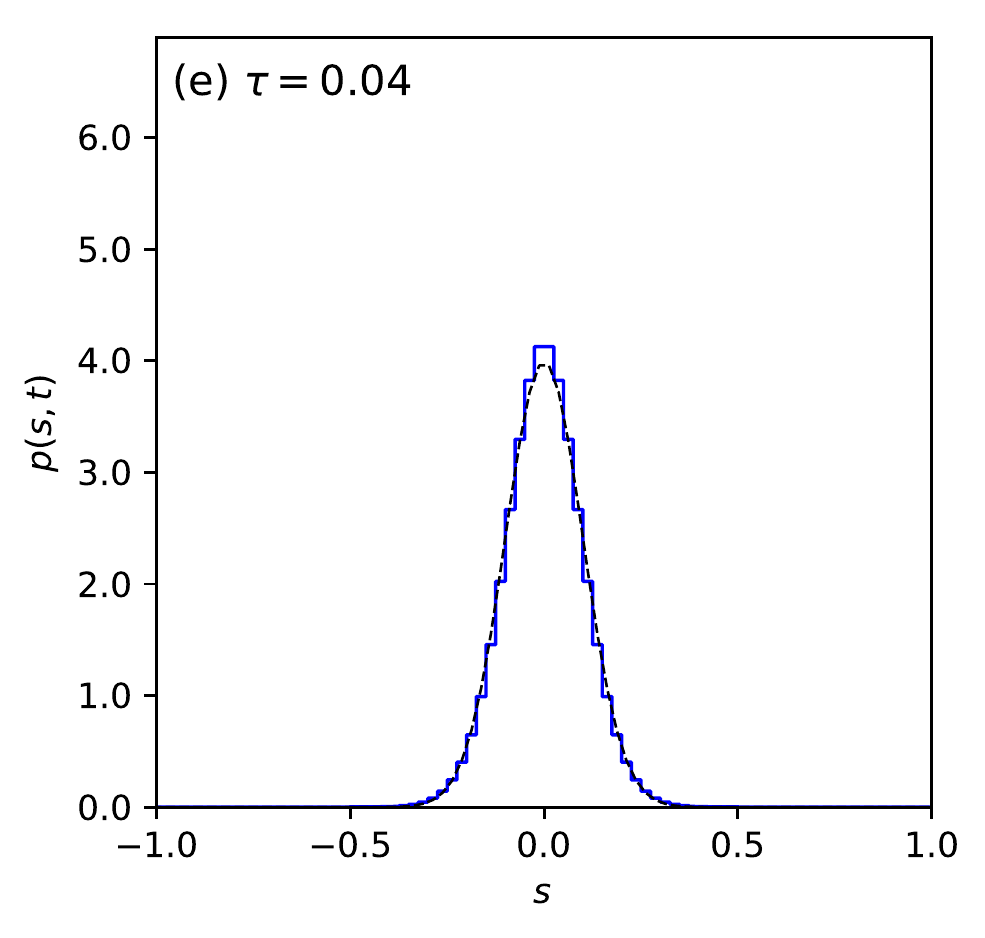}
  \includegraphics[width=0.33\textwidth]{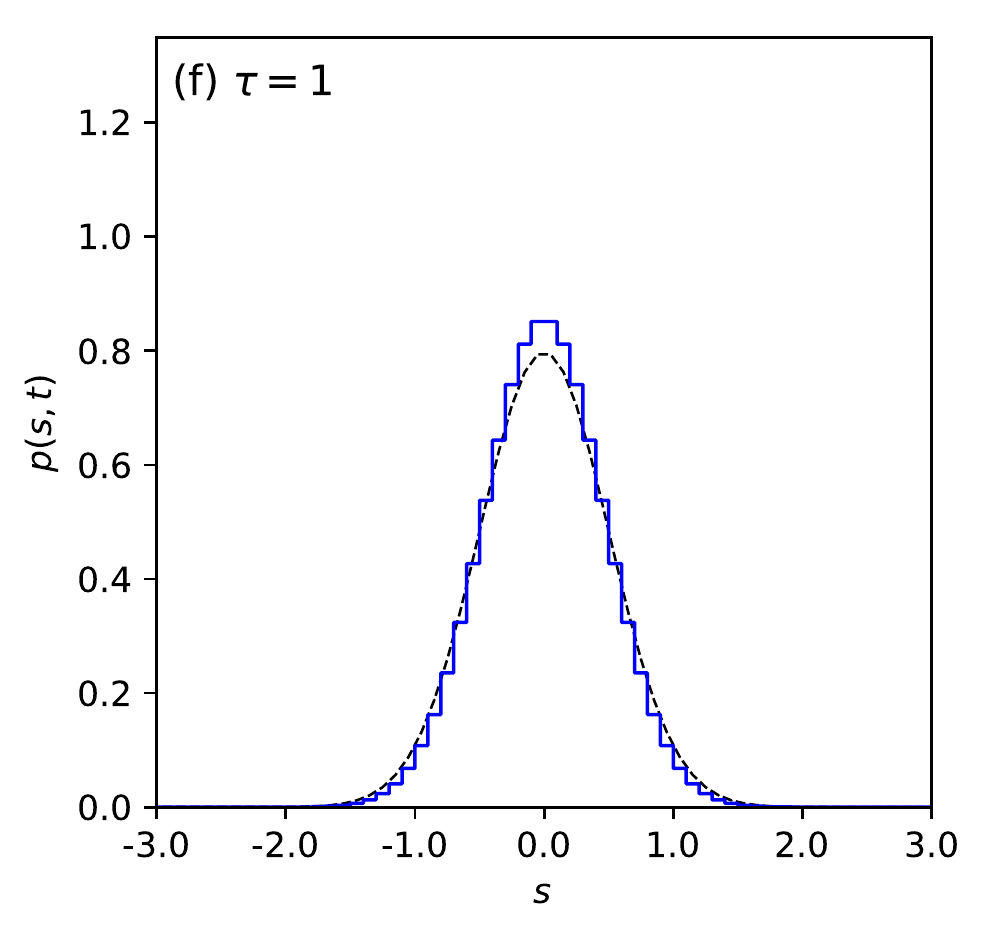}
}
\centerline{
  \includegraphics[width=0.33\textwidth]{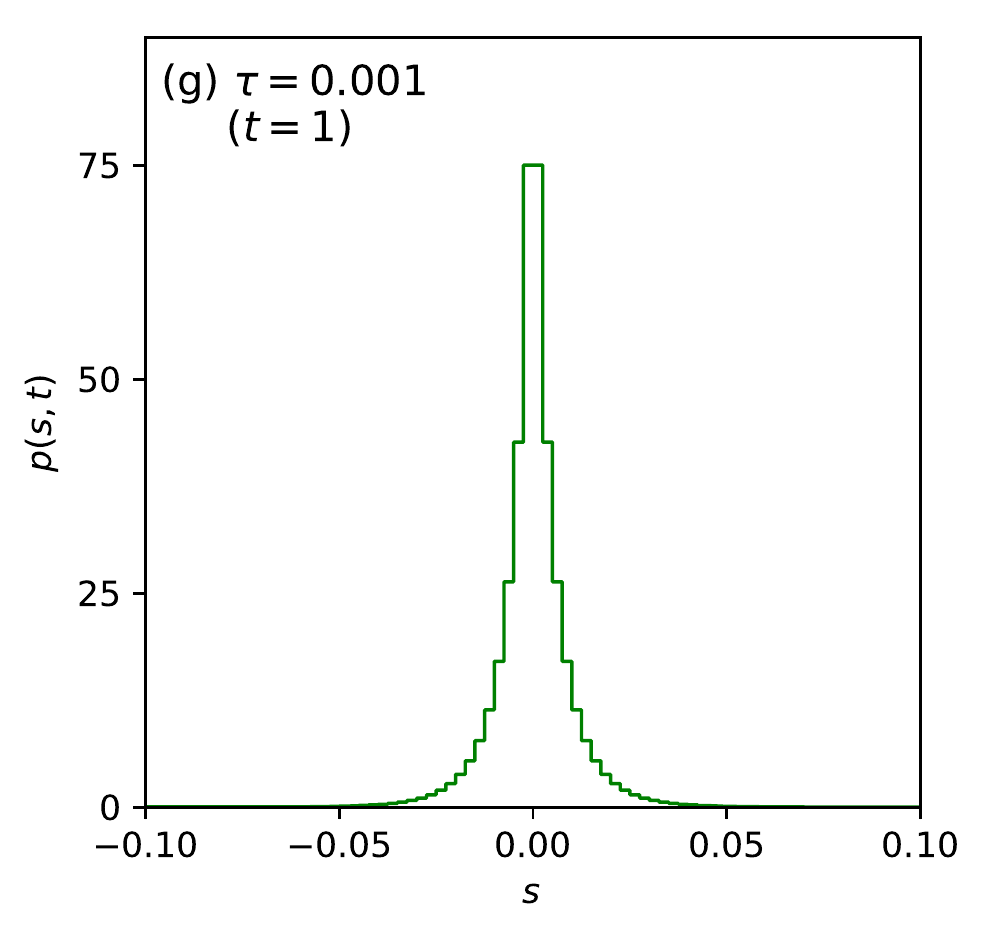}
  \includegraphics[width=0.33\textwidth]{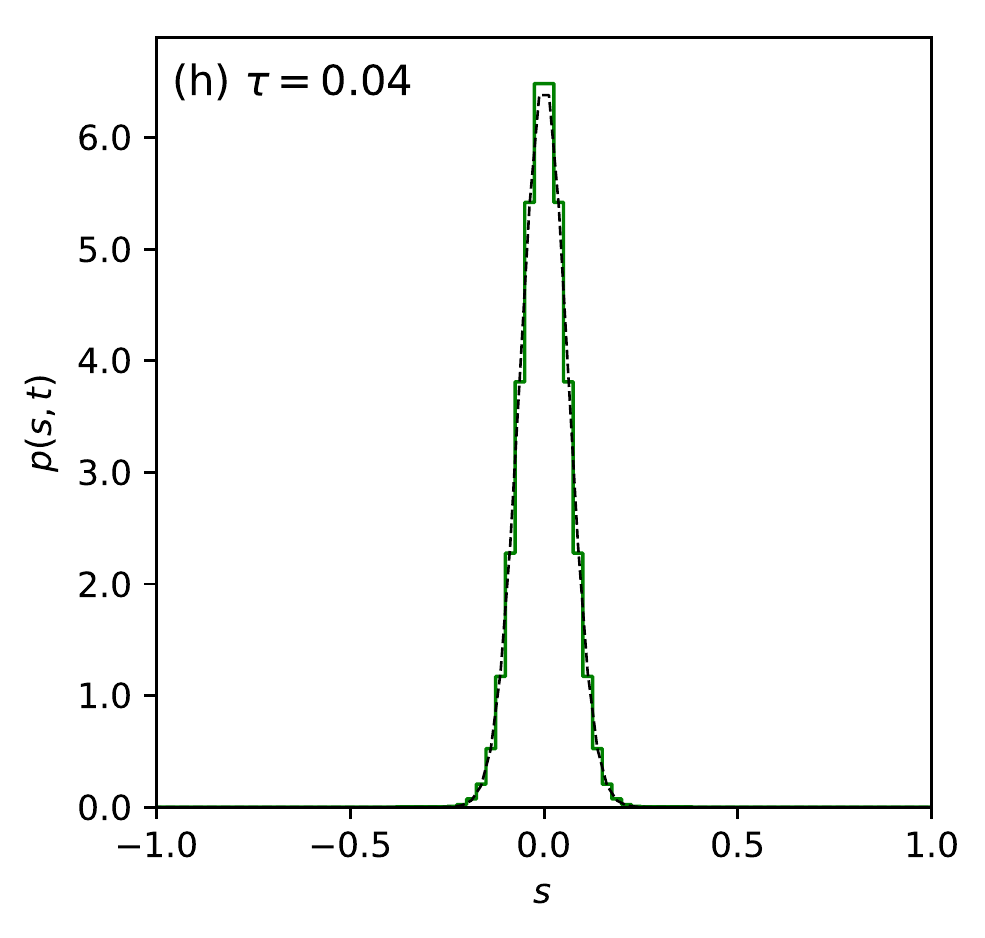}
  \includegraphics[width=0.33\textwidth]{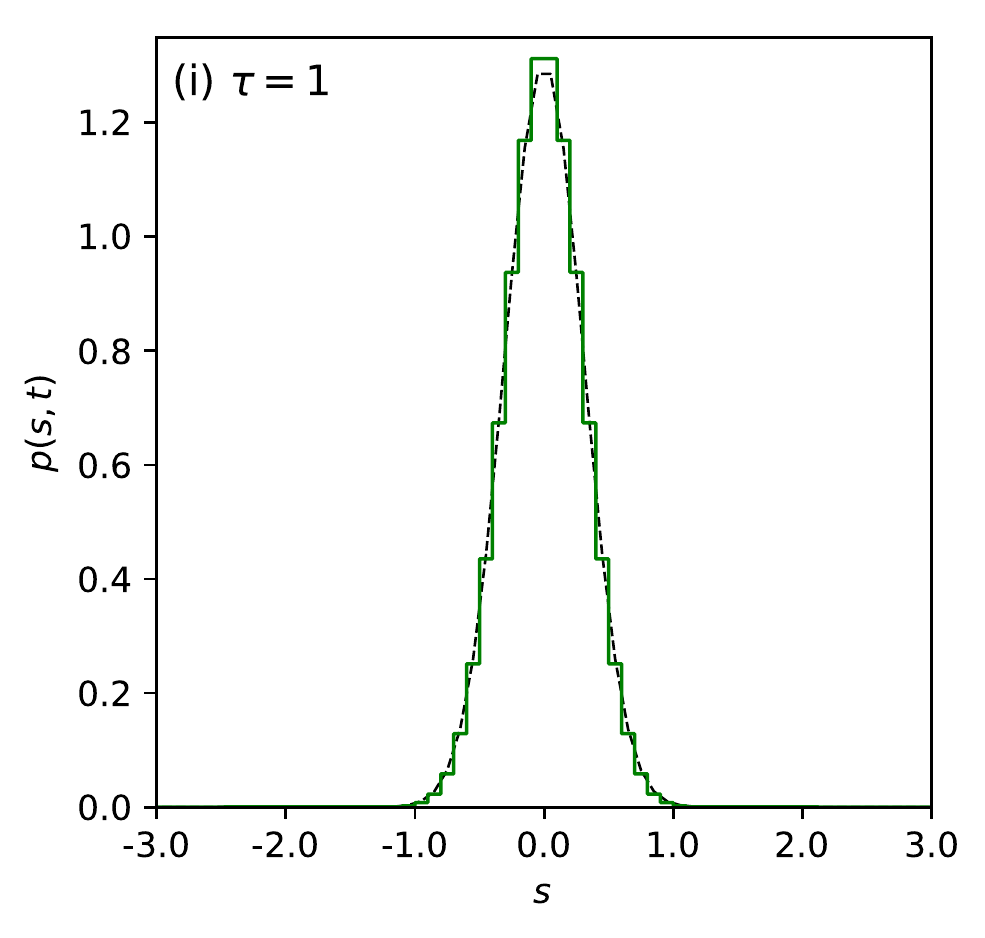}
}
\caption{\label{distLambda} For a partially randomized diagonal a histogram of
  the ensemble-averaged distribution of eigenvalue shifts $p(s,t)$ defined in
  Eq.~(\ref{pkxt}) is shown for GOE (top), GUE (middle) and GSE (bottom) with
  $N=1000$ and $t=1$ ($\tau=0.001$, left), $t=40$ ($\tau=0.04$, middle) and
  $t=1000$ ($\tau=1$, right).
  such that $s=1$ corresponds to the mean level spacing. The dashed lines in
  the middle and right columns show the theoretical distribution from
  Eq.~(\ref{pnx}). In (c) the full black line is a Gaussian fit to the data.}
\end{figure}

\begin{figure}[p]
\centerline{\includegraphics[width=0.8\textwidth]{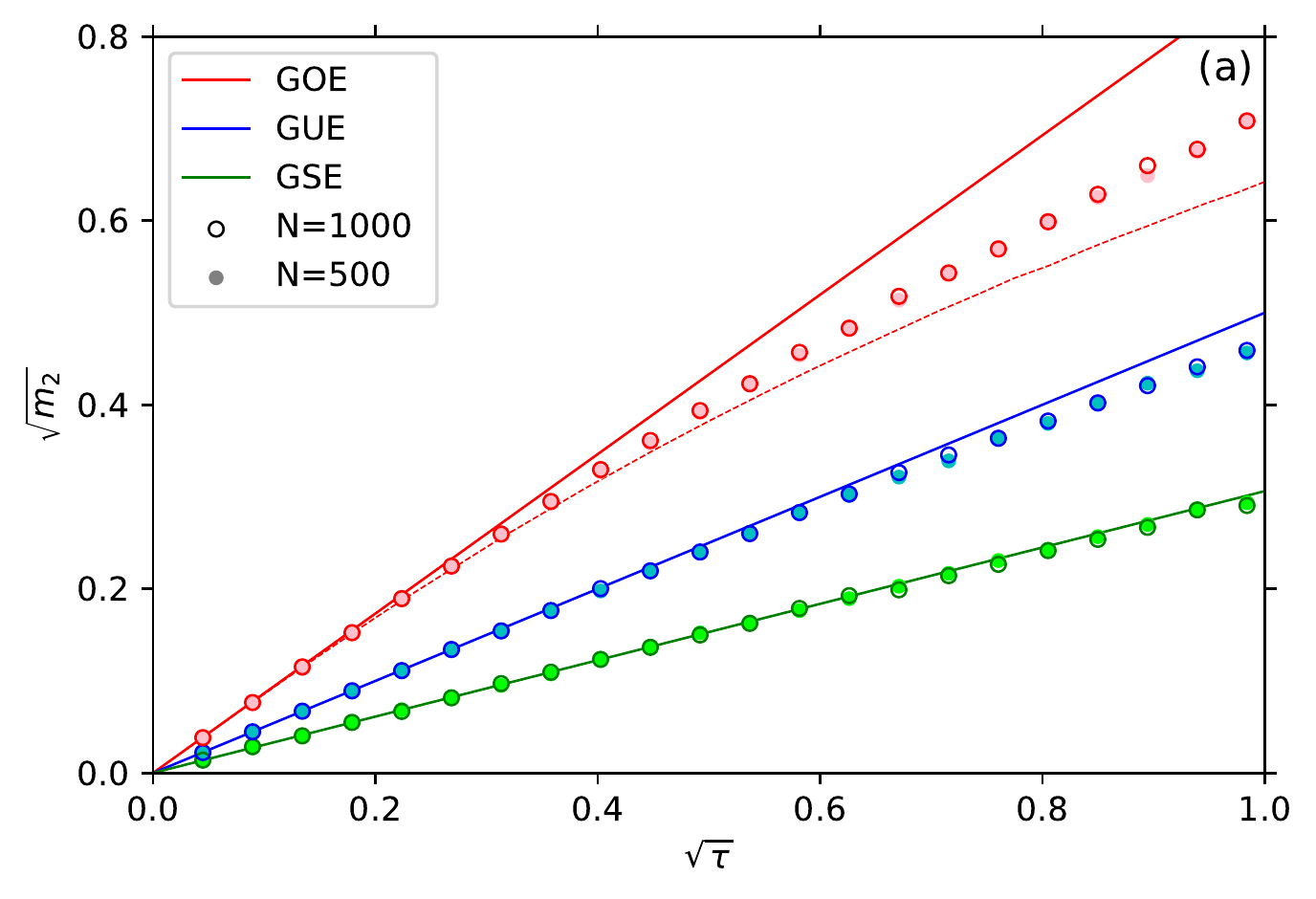}}
\centerline{\includegraphics[width=0.8\textwidth]{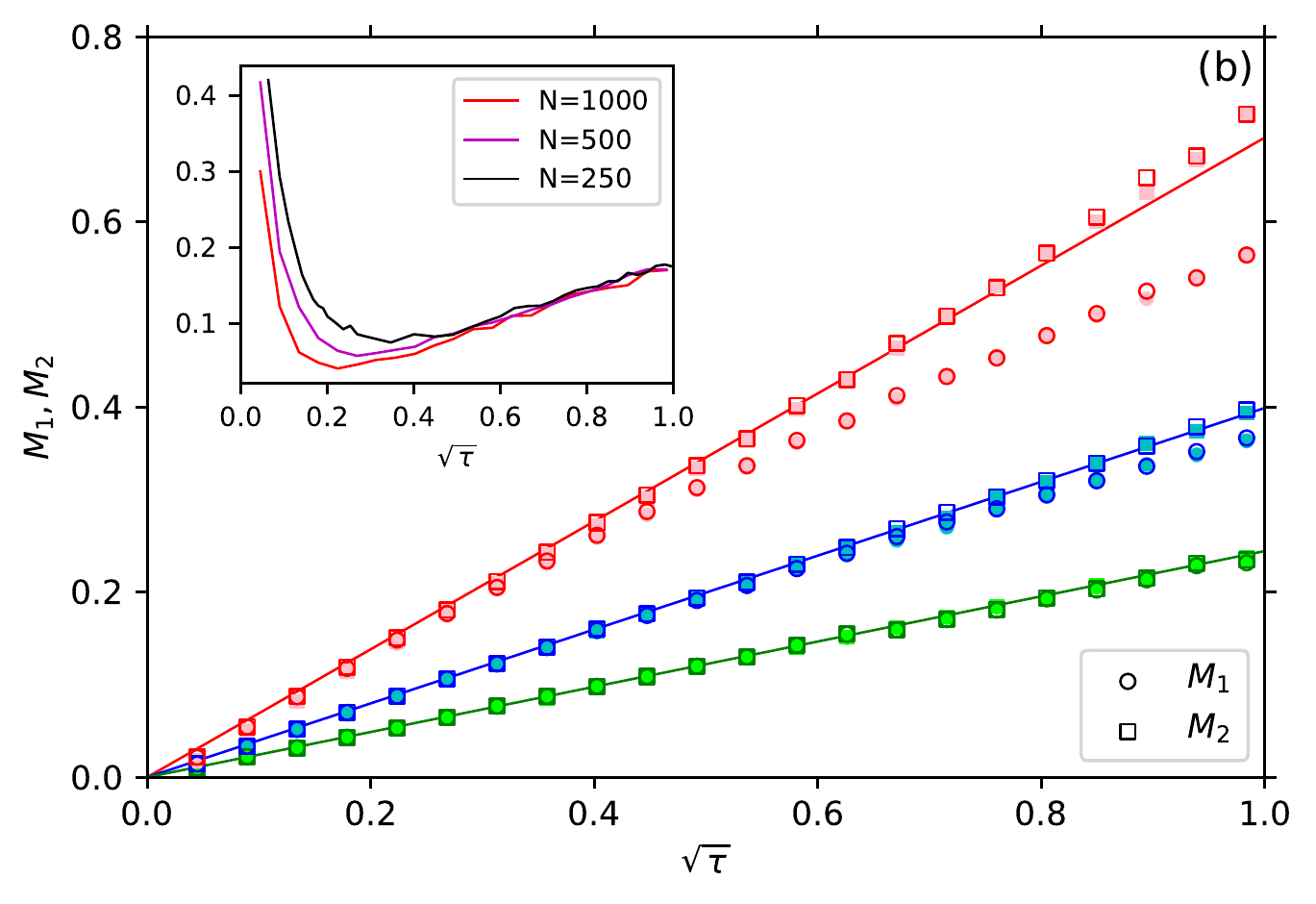}}
\caption{\label{randomized} Quantitative measures for the spectral variation
  for a partially randomized diagonal. \\ (a) Comparison of numerical results
  for the variance of individual eigenvalue shifts (dots) with results from
  Eq.~(\ref{m2}) (straight lines) for GOE (red), GUE (blue), and GSE (green).
  Note that the axes are $\sqrt{m_{2}}$ vs. $\sqrt{\tau}$. This scaling makes
  the horizontal axis compatible to that in Fig.~3b and maps our prediction
  on a straight line in order to highlight the nature of deviations.  The
  dashed line is a phenomenological correction of perturbation theory taking
  into account the re-ordering of the spectra when levels cross (see text).
  \\ (b) Numerical results for the first two moments of the spectral shift.
  The color code is the same as in Fig.~3a. The straight line is the
  perturbative result for the first moment given in Eq.~(\ref{M2rand}). The
  inset shows the relative deviation of $M_{1}$ from this prediction for the
  GOE. Additional colors are used here to display results for $N = 500$ and $N
  = 250$.}
\end{figure}

\begin{figure}
\centerline{
  \includegraphics[width=\textwidth]{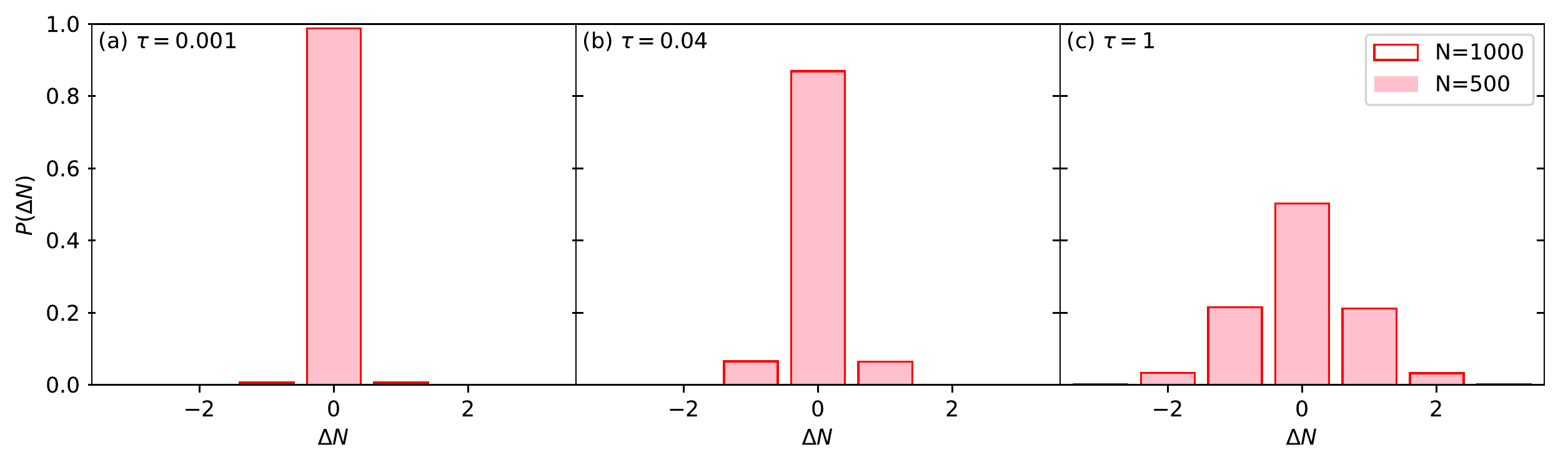}
}
\caption{\label{distShift} Probability distribution for the values of the
  spectral shift function for GOE for the same parameters as in Fig.~\ref{distLambda}a-c.}
\end{figure}

In this section we compare our theoretical predictions to numerical
simulations in order to verify and illustrate the results as well as to
discuss their range of applicability. All numerical ensemble averages are based on
  5000 realizations. For GSE ($\beta=4$) the standard representation of
  quaternions by $2\times2$ matrices was used which requires the
  diagonalization of a $2N\times2N$ matrix. It yields a doubly degenerate
  spectrum from which half of the eigenvalues were discarded.

  We first address in some detail the random replacement process.
  Fig.~\ref{distLambda} shows the ensemble averaged distribution of rescaled
  eigenvalue shifts $s_{n}=(\lambda_{n}^{(t)}-\lambda_{n}^{(0)})/\deltaAv$ for
  all three values of $\beta$. The selected numbers $t$ of randomized matrix
  elements correspond to three regimes of $\tau$ to be discussed below.
  Qualitatively we observe a distribution which is sharply peaked at zero when
  only few terms $t\sim1$ contribute in Eq.~(\ref {trandoms1st2}). It
  evolves into a Gaussian distribution as $t$ grows.  Note the varying scale
  on the horizontal axis. For small $\tau$ (left column) typical eigenvalue
  shifts are much smaller than the mean level spacing ($s\ll1$) while for
  maximal perturbation $\tau=1$ (right column) they are of the order of
  $\deltaAv$ ($s\sim1$). The dashed lines in the middle and right columns show
  our analytical result from Eq.~(\ref{pnx}). Agreement is best for GSE
  while for GOE there remain substantial deviations from our theory. Therefore
  we will concentrate on the GOE in the following and try to establish the
  range of applicability of our predictions.

Fig.~\ref{randomized}a shows the dependence on $\tau$ of the variance
$m_{2}^{(t)}$ of the distributions from Fig.~\ref{distLambda}. Note the
scaling $\sqrt{m_{2}}$ vs. $\sqrt{\tau}$ which was chosen for compatibility
with Fig.~\ref{randomized}b. The data for GOE are displayed with pink
and red color for two different matrix sizes $N=500$ and $N=1000$,
respectively.  Up to uncertainties due to the finite sample size, the results
are independent of $N$. Our analytical result (\ref{m2}) is plotted as
straight red line.  As we can see this prediction is valid for small values of
$\tau$ but overestimates the eigenvalue shifts for $\sqrt{\tau}\gtrsim0.2$. Besides the
gradual failure of perturbation theory for increasing $t$ this behavior can
also be attributed to level crossings which are not captured by our
perturbative treatment since the spectral measures we adopted are sensitive to
the ordering of eigenvalues.  Indeed, even if the perturbative results
$\hat\delta\lambda_{1}$ and $\hat\delta\lambda_{2}$ correctly describe the
shifts of two neighboring eigenvalues $\lambda_{1}<\lambda_{2}$ there can be
situations where
$\lambda_{1}+\hat\delta\lambda_{1}>\lambda_{2}+\hat\delta\lambda_{2}$,
i.e. the levels cross. This would imply that the perturbed spectrum is
  reordered such that $\lambda_{1}^{(t)}=\lambda_{2}+\hat\delta\lambda_{2}$ and
  $\lambda_{2}^{(t)}=\lambda_{1}+\hat\delta\lambda_{1}$. In this case the
  values for the eigenvalue shifts entering the calculation of $m_{2}^{(t)}$
  are smaller than the actual perturbative shifts. Such a failure of the
  perturbative approach is more likely for large perturbative eigenvalue
shifts, i.e. for growing $t$. It is also more likely if the probability of a
small level spacing is large. Therefore the deviation of our theory from the
data is considerably smaller for the GUE (Fig.~\ref{distLambda}d-f and blue
symbols in Fig.~\ref{randomized}a) where the level repulsion is stronger and
it is even smaller for GSE (green symbols).  To illustrate the effect of level
crossings we have calculated for GOE the perturbative spectrum numerically,
reordered it in an increasing sequence and then determined the distribution of
eigenvalue shifts.  The result for the associated second moment $m_{2}^{(t)}$
is plotted in Fig.~\ref{randomized}a as dashed line. It is closer to the
actual data but now underestimates the eigenvalue shifts. Thus, in order to get
a substantially better theoretical result it will be necessary to go beyond
first-order perturbation theory and to properly take into account the interaction between
the eigenvalues which becomes manifest, e.g. in avoided level crossings.
Yet, the essence of the behavior of $m_{2}^{(t)}$ for the various universality
classes is already contained in our analytical results.

\begin{figure}[p]
\centerline{\includegraphics[width=0.8\textwidth]{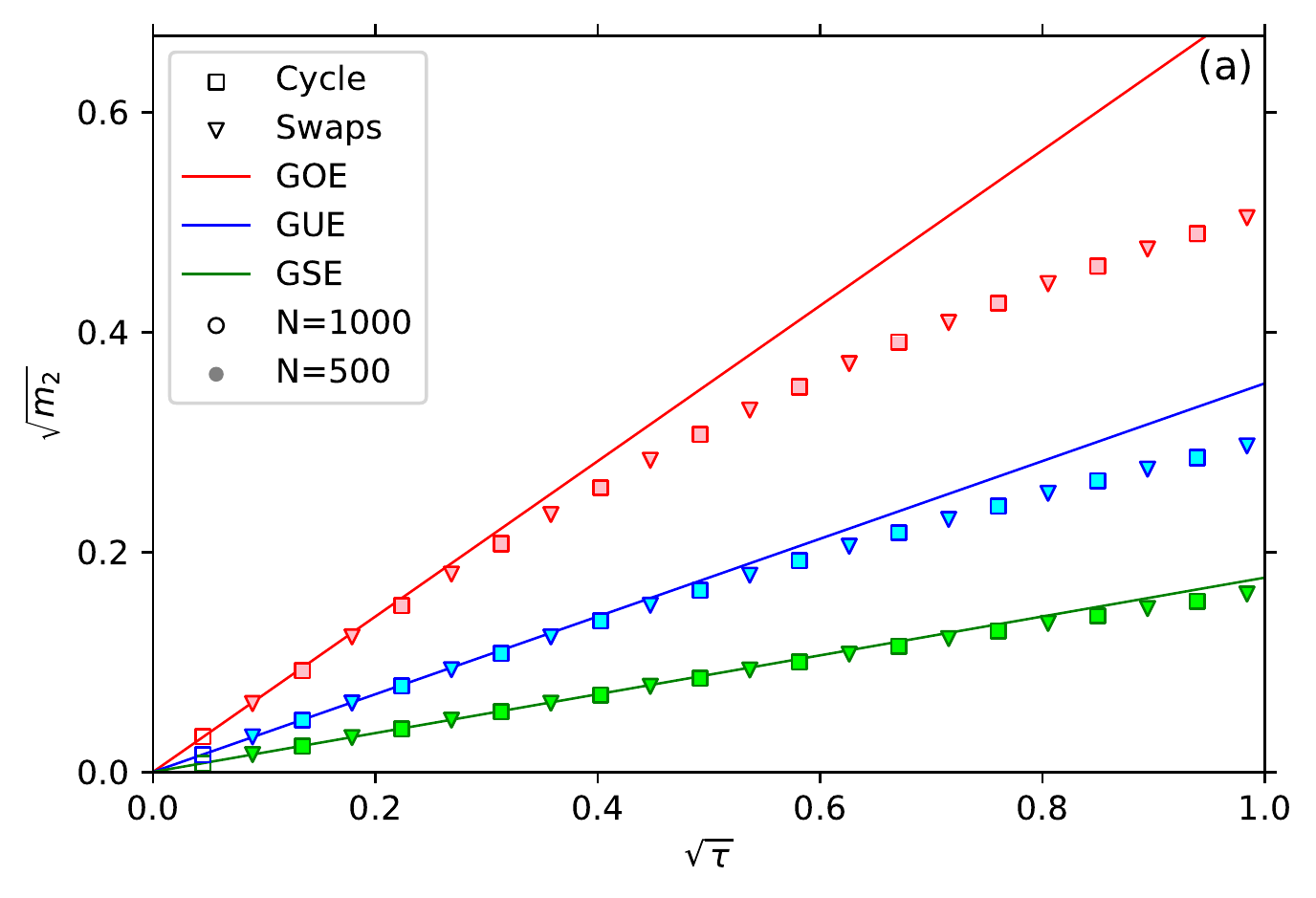}}
\centerline{\includegraphics[width=0.8\textwidth]{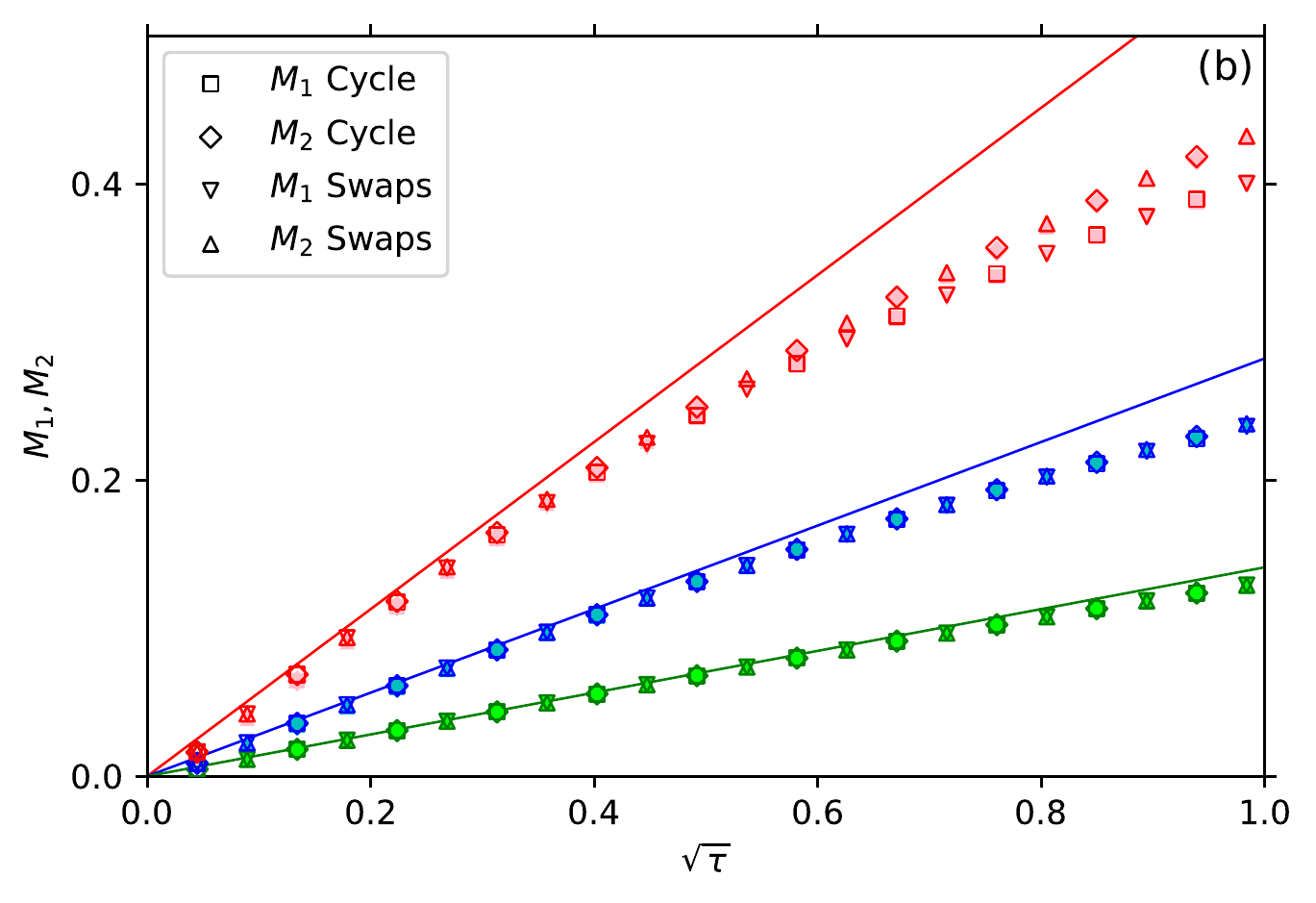}}
\caption{\label{permuted} Quantitative measures for the spectral variation for
  a partially permuted diagonal. The permutation is either cyclic, i.e. one
  cycle of length $t$ (squares and diamonds) or it consists of $t'=t/2$
  two-cycles, in which neighboring diagonal elements are swapped (triangles
  down and up). In both cases the minimum value of $t$ is 2 and in the latter
  case only even $t$ are allowed. \\ (a) Comparison of numerical results for
  the variance of individual eigenvalue shifts to the perturbative result
  (straight lines), see Eqs.~(\ref{m2p}), (\ref{cbp}). Note that the axes
  scaling are $\sqrt{\tau}$ vs. $\sqrt{m_{2}}$.  \\ (b) Numerical results for
  the first two moments of the spectral shift. The color code is the same as
  in (a). The straight lines show the perturbative result for the first
  moment, see Eqs.~(\ref{cbp}), (\ref{M2p}).  }
\end{figure}

In Fig.~\ref{randomized}b the first two moments $M_{1,2}$ of the spectral
shift are shown. According to Eq.~(\ref{m1}), $M_{1}=
m_{1}\sim\sqrt{m_{2}}$ and thus on first glance the data for $M_{1}$ (red
circles) show a behavior that is very similar to
Fig.~\ref{randomized}a. However, for small $\tau$ there are now deviations
which were absent in Fig.~\ref{randomized}a. They are magnified in the inset
of Fig.~\ref{randomized}b where the relative difference between the
simulations and our result $M_{1}^{(t)}=\sqrt{c_{\beta}\tau/2\pi}$ is
plotted. Indeed the proportionality between $\sqrt{m_{2}}$ and the first
moments $m_{1}=M_{1}$ relies on the validity of the Central Limit Theorem
which fails for small $t$. Consider the extreme case $t=1$
($\tau=N^{-1}$).  Taking the absolute value and summing over $n$ in
Eq.~(\ref{trandoms1st2}) yields
$\sum_{n=1}^{N}|\lambda_{n}^{(1)}-\lambda_{n}^{(0)}|=|\tilde
H_{1,1}-H_{1,1}|$. The r.h.s has mean value $\sqrt{2/\beta N\pi}$ such that
$M_{1}^{(1)}=\sqrt{1/2\beta N\pi}$ is the correct perturbative result for the
first moment. Its deviation from Eq.~(\ref{m1}),
$M_{1}^{(1)}=\sqrt{c_{\beta}/2N\pi}$, is about $42\%$ which is the maximal
value in the inset of Fig.~\ref{randomized}b. For increasing $\tau$ the
deviation decreases to a minimum before it grows again. For $N=1000$ this minimum
corresponds to the distribution shown in Fig.~\ref{distLambda}b
($\tau=0.04$). It is a compromise between the error in the Central Limit
Theorem for small $t$ and the error in perturbation theory for large $t$. The
minimal deviation decreases as $N$ increases and at the same time it is reached
for smaller values of $\tau$ such that Eq.~(\ref{m1}) is valid with increasing
precision and in a growing range.

As explained below Eq.~(\ref{Mt}), the first two moments of the spectral shift
are related by $M_{1}^{(t)}\le M_{2}^{(t)}$ with equality holding when the
distribution is limited to values $|r|\le 1$. This is confirmed by
Figs.\ \ref{randomized}b and \ref{distShift}. For small $\tau$ the average
eigenvalue shift is much smaller than the mean level spacing (Fig.~\ref{distLambda}a,b)
and therefore the spectral shift typically does not exceed one
(Fig.~\ref{distShift}a,b). In Fig.~\ref{randomized}b we observe that the data
for $M_{1,2}^{(t)}$ coincide in this regime. Beyond that range the data for
$M_{2}^{(t)}$ grow faster and happen to be quite close to the perturbative
prediction for $M_{1}^{(t)}\!\!,$ but we have no quantitative explanation for
this observation.

Fig.~\ref{permuted} shows $m_{2}^{(t)}$ and $M_{1,2}^{(t)}$ for the two types
of permutations of diagonal matrix elements which were discussed in detail in
Section \ref{processI}. Squares represent a single cycle of length $t$ while
triangles represent $t/2$ cycles of length 2. As expected, both types of
permutations lead to similar results which for small $\tau$ approach our
predictions. The qualitative behavior is very similar to randomized diagonal
elements (Fig.~\ref{randomized}) but the relative deviation from our theory is
slightly larger. In Fig.~\ref{permuted}b there is again a minimal deviation
for $M_{1}^{(t)}$ at some intermediate value of $\tau$ compromising between
perturbation theory and Central Limit Theorem which are valid for small $t$
and large $t$, respectively. In contrast to Fig.~\ref{randomized}, the
difference between $M_{1}^{(t)}$ and $M_2^{(t)}$ is smaller now such that for
$\tau\sim1$ the second moment of the spectral shift remains well below our
perturbative result for $M_{1}^{(t)}$.

The behavior of GOE, GUE and GSE is qualitatively very similar in both,
Fig.~\ref{randomized} and Fig.~\ref{permuted}. The only difference is in the
magnitude of the deviations from our theory which is decreasing with
increasing $\beta$. We attribute this to the increasing level repulsion from 
GOE to GSE which implies a lower probability of small level
spacings. Thus, also perturbative level crossings (or avoided crossings of
levels under the perturbation), which were identified above as an important
source of deviations, have less influence.

\section{Random replacement of columns and rows}

The process \emph{(iii)} defined in the Introduction differs considerably from
the two processes \emph{(i)} and \emph{(ii)} discussed previously: once it is
completed at $t=N$ the resulting matrix is completely independent of the
reference matrix, in stark contrast to processes \emph{(i)} and \emph{(ii)}
where only the diagonal matrix elements are affected by the perturbation.

We shall start this section by explaining why the perturbative approach of
Section \ref{diagPerturb} fails for process \emph{(iii)}. Using
Eq.~(\ref{trandomhans}) first-order perturbation theory yields
\begin{eqnarray}
\hspace {-20mm}
\lambda_n^{(t)}-\lambda_n^{(0)}&=&\sum_{\mu=1}^t(\tilde H_{\mu,\mu }-H_{\mu,\mu})\ |\left\langle n |\mu\right\rangle|^2 \\
&+&
\sum_{\mu=1}^t\sum_{\nu>\mu}^N\left [ (\tilde H_{\mu,\nu }-H_{\mu,\nu})\ \left \langle n |\mu\rangle\langle \nu|n \right \rangle+  (\tilde H_{\nu,\mu}-H_{\nu,\mu})\ \left \langle n|\nu\rangle\langle \mu|n \right \rangle \right ] ,\nonumber
\end{eqnarray}
where $\tilde H$ is a randomly chosen matrix from the respective ensemble. The
sums differ in the number of their summands, and in their variances.  Each of
them is composed of independent and identically distributed random
numbers. Thus, in the limit of large $N$ and $t$ we can apply for each sum the
Central Limit Theorem. The first sum coincides with Eq.~(\ref{trandoms1st2})
and yields a level shift of the order of the mean level spacing or below, as
shown in Section \ref{processII}. Furthermore, the scaling with $N$ is of the
same order of magnitude for each summand of both sums. However, the number of
terms in the second sum is larger by a factor $\sim N$. Thus, this sum
compared to the mean level spacing is larger by a factor growing to infinity
$\sim\sqrt{N}$. Accordingly, first-order perturbation theory fails.

\begin{figure}[t]
  \centerline{
    \includegraphics[height=0.48\textwidth]{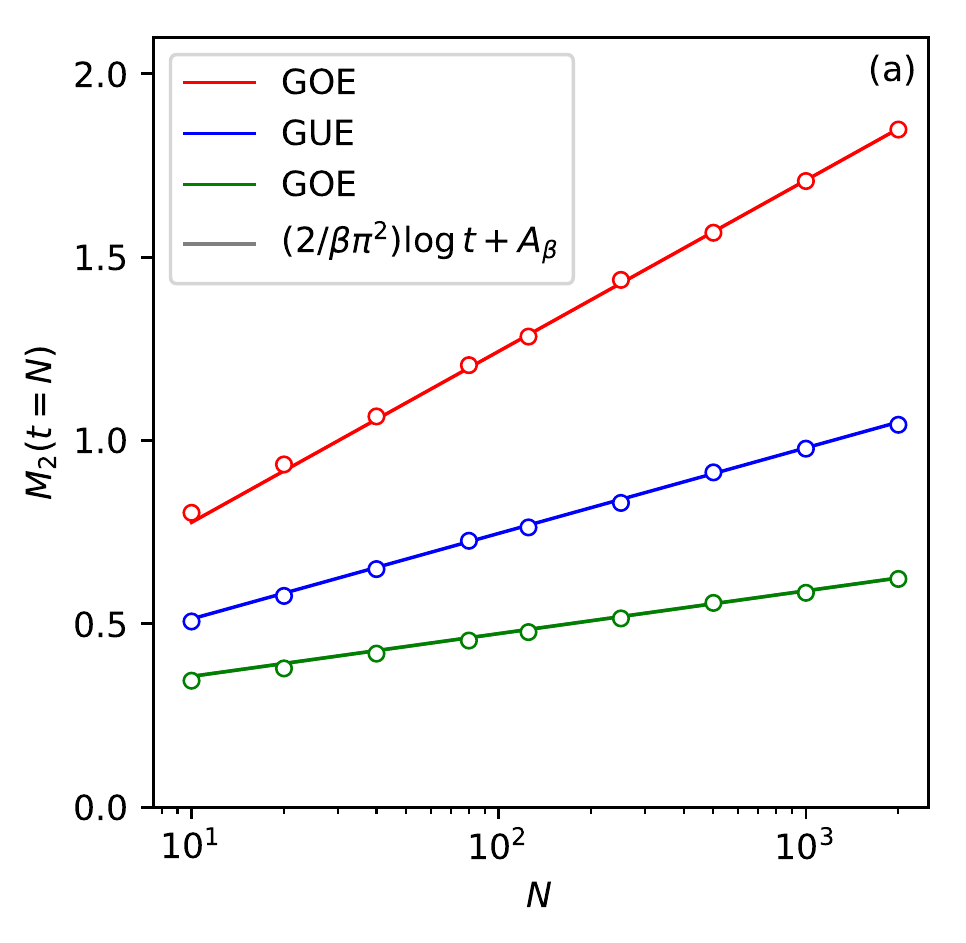}
    \includegraphics[height=0.48\textwidth]{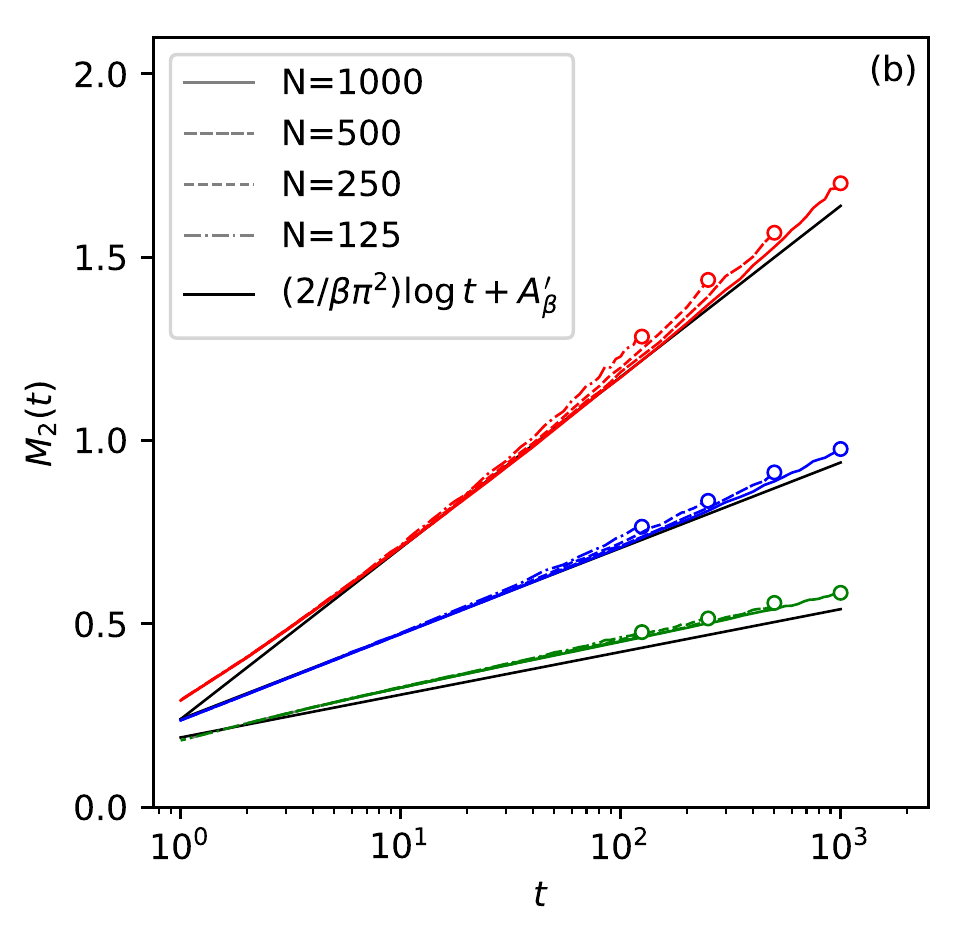}
  }
  \centerline{
    \includegraphics[height=0.36\textwidth]{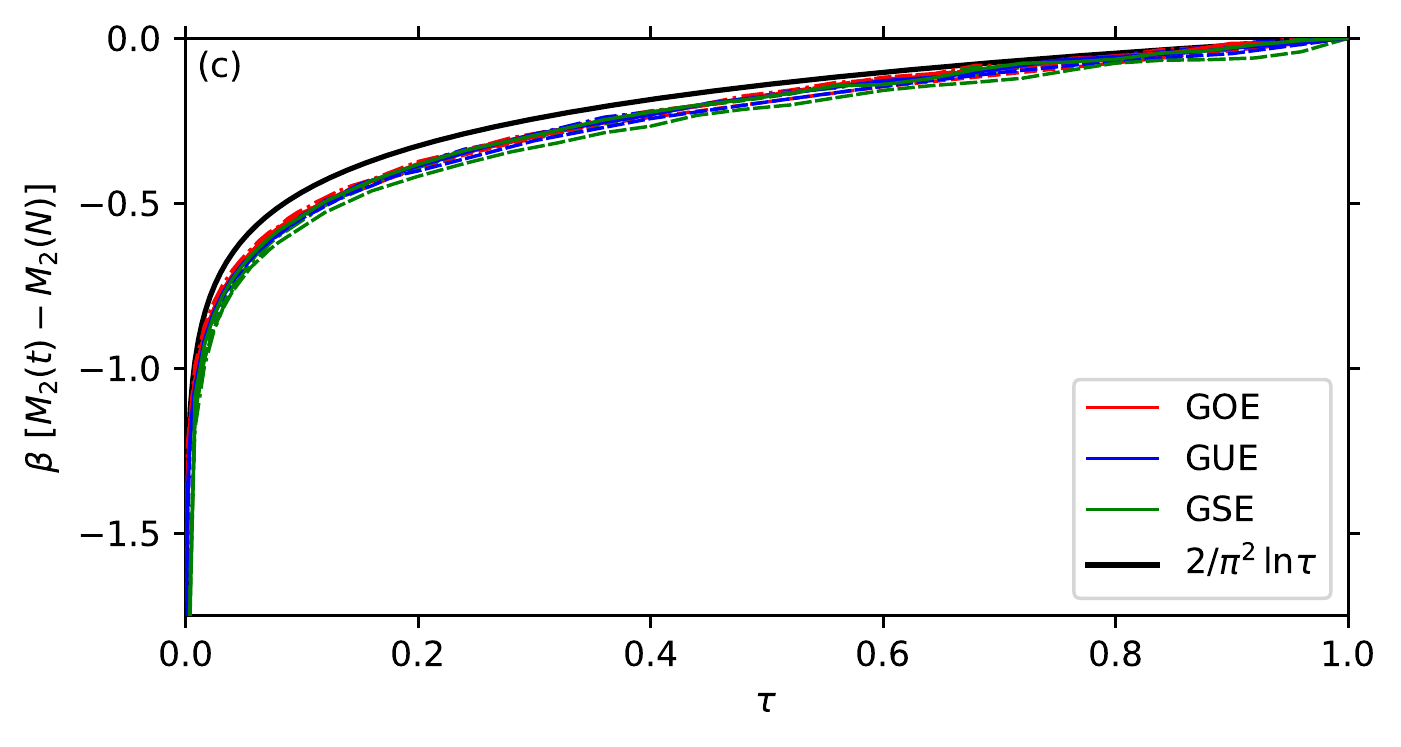}
  }
  \caption{\label{MHW} Variance of the spectral shift for a perturbation
    where $t$ rows and columns are replaced by random values. (a) shows the
    dependence on $N$ of the variance for completely uncorrelated matrices
    ($t=N$) with open dots and compares it to the logarithmic increase
    predicted in Eq.~(\ref{M2t=N}).  (b) shows the dependence on $t$ for some
    selected values of $N$ by continuous lines (color code as in (a)). The end
    points of these curves at $t=N$ are marked by open dots as in (a). The
    black lines correspond to Eq.~(\ref{M2t}).  (c) shows the same data as (b)
    as a function of $\tau=t/N$. The value at $t=N$ has been subtracted such
    that all curves are zero at $\tau=1$ and they are scaled by a factor
    $\beta$. The line styles are chosen as in (b) to distinguish various
    values of $N$. However, with the adopted scaling all curves fall on top of
    each other within the numerical accuracy. The black line corresponds to
    Eq.~(\ref{scaling}).  }
\end{figure}

So far we cannot offer a complete theory for the process
\emph{(iii)}. However, we will present numerical and analytical approaches towards
an understanding of the scaling of the spectral shift with the parameters $N$,
$t$ and $\beta$.

The fact that the matrices $H(t=N)$ and the reference matrix $H(t=0)$ are
statistically independent can be used to establish constraints on the variance
$M_2$. In Eq.~(\ref{meanN}) the average spectral counting function $\left
\langle \mathcal{N} (\lambda ) \right \rangle $\ is introduced which is
independent of $t$. We consider the fluctuations of the actual spectral
counting function around $\left \langle \mathcal{N} (\lambda ) \right \rangle
$, $\delta \mathcal{N} (\lambda;t) = \mathcal{N} (\lambda;t)- \left \langle
\mathcal{N} (\lambda )\right \rangle $. Then
\begin{eqnarray}
\hspace{-25mm} M_{2}^{(t)} &=& \frac{ 1}{2}\int_{-1}^{1} d\lambda \left \langle
\left[ \mathcal{N} (\lambda;t)-\mathcal{N} (\lambda;0)\right ]^2 \right
\rangle = \frac{ 1}{2}\int_{-1}^{1} d\lambda \left \langle \left[ \delta
  \mathcal{N} (\lambda;t)-\delta \mathcal{N} (\lambda;0)\right ]^2 \right
\rangle \\
\hspace{-25mm}
 M_{2}^{(N)} &=& \frac{ 1}{2}\left [ \int_{-1}^{1} d\lambda  \langle  [ \delta \mathcal{N} (\lambda;N)]^2  \rangle 
+ \int_{-1}^{1} d\lambda  \langle \left [  \delta \mathcal{N} (\lambda;0)\right ]^2 \rangle \right ]  
-  \int_{-1}^{1} d\lambda \left \langle    \delta \mathcal{N} (\lambda;N) \delta  \mathcal{N} (\lambda;0) \right \rangle  \nonumber      
\end{eqnarray}

The two integrals bracketed in the square parentheses correspond to leading order in $N$ to the number variance $\Sigma^2(L=N)$ introduced by Dyson \cite{Dyson} which, in the limit of large $N$, is well approximated by~\cite{Meh90}
\begin{equation}\label{sig2}
\Sigma^2(N) = \frac{2}{\beta \pi^2}\log N + a_{\beta}
\end{equation}
where $a_{\beta}$ is independent of $N$.
The third integral is zero due to the absence of correlations between
$\delta\mathcal{N}(\lambda;0)$ and $\delta\mathcal{N}(\lambda;N)$.
Thus we expect that the variance for $t=N$ grows logarithmically
with $N$,  
\begin{eqnarray}\label{M2t=N}
M_{2}^{(N)}&=&\frac{2}{\beta \pi^2}\log N + A_{\beta}\,.
\end{eqnarray}
Indeed this is confirmed in Fig.~\ref{MHW}a.
The constant $a_\beta$ is known explicitly in random matrix theory. There
the eigenvalues, the spectral density of which is given by the semicircle law,
are unfolded to uniform density. We, however, use raw spectra for the
computation of $M_{2}^{(N)}$. Accordingly, we have determined $A_{\beta}$ by
fitting Eq.~(\ref{M2t=N}) to the numerical curves.

In Fig.~\ref{MHW}b the dependence of $M_{2}^{(t)}$ on $t$ is shown for $N=$
125, 250, 500 and 1000. A remarkable numerical observation is that these
curves all fall on top of each other when $t$ is not too close to its maximum
value $t=N$. For each ensemble a logarithmic increase is observed
with a slope and a constant close to the values from (\ref{M2t=N}), i.e. we have
\begin{equation}\label{M2t}
  M_{2}^{(t)}=\frac{2}{\beta \pi^2}\log t + A_{\beta}'\,.
\end{equation}
The numerical data follow this law for a range of $t$ which is growing with
$N$. Note however, that the case $t=0$ with $M_{2}^{(0)}=0$ is not shown in
Fig.~\ref{MHW}b because of the logarithmic scale.

Finally, in Fig.~\ref{MHW}c we show the data from Fig.~\ref{MHW}b in a
different representaion where the endpoints of all curves coincide. Namely we
consider the quantity $M_{2}^{(t)}-M_{2}^{(N)}$. According to Eq.~(\ref{M2t})
it can be represented as a function of $\tau=t/N$ and is inversely
proportional to $\beta$,
\begin{equation}\label{scaling}
M_{2}^{(t)}-M_{2}^{(N)}=\frac{2}{\beta \pi^2}(\log t-\log N)=\frac{2}{\beta \pi^2}\log\tau\,.
\end{equation}
Indeed, when scaled by $\beta$, the curves for all matrix sizes $N$ and all
ensembles essentially fall on top of each other as shown in
Fig.~\ref{MHW}c. An exception is the case $\tau=0$ where a logarithmic singularity
$\sim-\log N$ develops.

Thus, we may conclude that a perturbation randomizing $t$ rows and the
corresponding $t$ columns of the reference matrix $H^{(0)}$ leads to
$M_{2}^{(t)}\simeq\Sigma^2(t)$ plus some lower-order corrections including a
constant term $A_{\beta}'-A_{\beta}$.
Further work is needed to obtain a more accurate description of these
corrections. It could be based, e.g., on the supersymmetric approach to
random-matrix theory \cite{VWZ85,GW90,Guh91,JSW17}.

\section{Conclusions}\label{conclusions}

We have studied the effect of perturbations of rank $t$ on the spectra of
Gaussian random matrices. Among other spectral measures we considered the
variance of the spectral shift function. From the virtually unlimited number
of possibilities to define perturbations of a given rank we chose two quite disparate
cases, one where only diagonal elements are affected and one where entire rows
and columns are randomized. In the first case we obtain a satisfactory
theory using first-order perturbation theory. In the second case perturbation
theory fails and we have no complete analytical theory at the moment. However, we
established numerically the scaling of the spectral shift with the
parameters $N$ and $t$. While in the first case all spectral measures
considered can be expressed in terms of scaled variable $\tau= \frac{t}{N}$
alone, in the second case there is an additional dependence on the matrix
size $N$.

It is definitely possible to extend our results beyond the considered
perturbation models. For example, if we randomize a quadratic block of size
$t$ we have a case which interpolates between the two cases we studied and we
observe a crossover between perturbation theory for small $t$ and the case
of two completely uncorrelated matrices at $t=N$. 

One central result for the case of a diagonal perturbation is that the
variance of the spectral shift increases as $D \tau^{\frac{1}{2}}$
(Eqs.~(\ref{M2rand}) and (\ref{M2p})) with some constant $D$. This holds for
$N\gg1$ in a parameter range where $t\gg1$ is large but $\tau=t/N\ll1$ is
small.  The numerically observed range of validity is increasing with $\beta$
so that it covers the entire range of $\tau$ for the GSE, is slightly
confined for GUE and is limited to small $\tau$ for GOE.

An interesting interpretation of this result arises if we consider $t$ as time
and the gradually growing spectral variation as a dynamic process. At each
time step $t\to t+1$ there is a slight variation of the spectrum, e.g. by
randomizing another diagonal element. These steps result in a random walk in
the spectral shift distribution where the resulting mean
probability distribution displays anomalous diffusion (subdiffusion)
in the "time" $\tau$.  The "diffusion coefficient" $D$ depends on the type of
perturbation and the random matrix ensemble under study. This
behavior can be compared to the spectral shift statistics of a different
model, namely the Anderson model in one dimension which corresponds to an
ensemble of Jacobi matrices with random \emph{iid} diagonal elements and $1$
on the secondary diagonals. Heuristic arguments supported by some estimates
suggest that in this case the variance of the spectral shift due to
interchanging or randomizing $t$ diagonal elements grows linearly with
$\tau$. Numerical simulations confirm this intuition \cite{workinprogress}. In
other words, the Poissonian nature of the spectrum seems to lead to standard
diffusion of the spectral shift.

Anomalous diffusion with variance which grows in time as $D\tau
^{\frac{1}{2}}$ occurs in several systems which were studied in statistical
mechanics out of equilibrium. Particularly similar to the system
under study in the present article is the single file tracer model, in which particles are
confined to a tube and scatter elastically when they get closer, but are
unable to penetrate each other.
The displacements of tracer
particles are known to be sub-diffusive with the variance proportional to the
square root of the time. It was recently shown \cite{Miron19} that a
transition to regular diffusion, with a variance growing linearly in time, occurs
once particles are allowed to interchange their position. The question whether
this similarity to our problem is rooted in the dynamics of the two systems, or it is just
incidental, remains to be addressed.

\section *{Acknowledgements}
Michael Aizenman and Jonathan Breuer introduced US to the martingale concept
and its application in the present case. Thanks to both. In particular we are
obliged to Jonathan who provided us with a lemma on $S_{t}$ and its proof,
used in the discussion of equations (41-44). We thank David Mukamel for
pointing out the possible connection between the single file tracer model and
the present work. B.D. thanks the NNSF of China for financial support under
grant Nos. 11775100 and 11961131009 and the Weizmann Institute of Science for
financial support and hospitality.

\end{document}